\documentclass[aps,prd,nofootinbib,amsmath,amssymb,showpacs,longbibliography,singlecolumn,superscriptaddress,notitlepage]{revtex4-1}

\usepackage{epsfig}
\usepackage{bm}
\usepackage{amssymb}
\usepackage{amsmath}
\usepackage{dsfont}
\usepackage{color}
\usepackage{subfigure}
\usepackage{dcolumn}
\usepackage{mathtools}
\usepackage{enumerate}
\usepackage[utf8]{inputenc}
\usepackage{soul}
\usepackage{float}

\usepackage{amsmath}
\usepackage{amssymb}
\usepackage{amsthm}
\usepackage{amscd, amsfonts, mathrsfs}
\usepackage{cases}
\usepackage{verbatim} 
\usepackage{epsf}
\usepackage[bookmarksnumbered,pdfpagelabels=true,plainpages=false,colorlinks=true,
            linkcolor=black,citecolor=black,urlcolor=black]{hyperref}

\theoremstyle{plain}

\theoremstyle{definition}

\allowdisplaybreaks

\makeatletter
\newcommand{\unitaryi}[2]{\ensuremath{U_{(#1,#2)}}}
\newcommand{\unitarydaggi}[2]{\ensuremath{U_{(#1,#2)}}}
\newcommand{\unitaryinsertfield}[1]{\ensuremath{[#1]}}
\newcommand{\@unitaryii}{\@ifnextchar\bgroup{\unitaryinsertfield}{}}
\newcommand{\unitary}[2]{\unitaryi{#1}{#2}\@unitaryii}
\newcommand{\unitarydagg}[2]{\unitarydaggi{#1}{#2}\@unitaryii}
\makeatother

\newcommand{\tr}{\operatorname{tr}}

\newcommand{\be}{\begin{equation}}
\newcommand{\ee}{\end{equation}}
\newcommand{\bea}{\begin{eqnarray}}
\newcommand{\eea}{\end{eqnarray}}

\newcommand{\bml}{\begin{subequations}}
\newcommand{\eml}{\end{subequations}}

\newcommand{\bbm}{\begin{bmatrix}}
\newcommand{\ebm}{\end{bmatrix}}
\newcommand{\bvm}{\begin{vmatrix}}
\newcommand{\evm}{\end{vmatrix}}

\begin{document}

\title{Emergent Viscous Hydrodynamics From a Single Quantum Particle} 

\author{Zhi-Li {Zhou}}
\email{zhiliz2@illinois.edu}
\affiliation{Department of Physics and Illinois Center for Advanced Studies of the Universe, University of Illinois Urbana-Champaign, 1110 West Green Street, Urbana, Illinois 61801, USA}

\author{Mauricio {Hippert}}
\email{hippert@cbpf.br}
\affiliation{Centro Brasileiro de Pesquisas Físicas, Rua Dr. Xavier Sigaud 150, 
Rio de Janeiro, RJ, 22290-180, Brazil}
\affiliation{Instituto de Física, Universidade do Estado do Rio de Janeiro, Rua São Francisco Xavier, 524, Rio de Janeiro, RJ, 20550-013, Brazil}

\author{Nicki {Mullins}}
\email{nickim2@illinois.edu}
\affiliation{Department of Physics and Illinois Center for Advanced Studies of the Universe, University of Illinois Urbana-Champaign, 1110 West Green Street, Urbana, Illinois 61801, USA}

\author{Jorge {Noronha}}
\email{jn0508@illinois.edu}
\affiliation{Department of Physics and Illinois Center for Advanced Studies of the Universe, University of Illinois Urbana-Champaign, 1110 West Green Street, Urbana, Illinois 61801, USA}


\begin{abstract}
We investigate an explicit example of how spatial decoherence can lead to hydrodynamic behavior in the late-time, long-wavelength regime of open quantum systems. We focus on the case of a single non-relativistic quantum particle linearly coupled to a thermal bath of noninteracting harmonic oscillators at temperature $T$, \emph{a la} Caldeira and Leggett. Taking advantage of decoherence in the position representation, we expand the reduced density matrix in powers of the off-diagonal spatial components, so that high-order terms are suppressed at late times. Truncating the resulting power series at second order leads to a set of dissipative transient hydrodynamic equations similar to the non-relativistic limit of equations widely used in simulations of the quark-gluon plasma formed in ultrarelativistic heavy-ion collisions. Transport coefficients are directly determined by the damping constant $\gamma$, which quantifies the influence of the environment. 
The asymptotic limit of our hydrodynamic equations reduces to the celebrated Navier-Stokes equations for a compressible fluid in the presence of a drag force. Our results shed new light on the onset of hydrodynamic behavior in open quantum systems where a system with few degrees of freedom is coupled to a large thermal environment.

\end{abstract}




\maketitle

\tableofcontents

\section{Introduction} \label{sec:introduction}

There is much work on emergent hydrodynamics in quantum many-body systems (e.g., see \cite{ PhysRevX.6.041065,PhysRevX.8.021013, PhysRevLett.125.030601,Zu_2021}) and how to use hydrodynamics to describe quantum plasmas (for instance, see \cite{haas2011quantum, Michta_2015, 10.1063/1.5097885}). 
Interest in the subject is not limited to non-relativistic quantum systems, as hydrodynamics is strikingly effective in the description of the quark-gluon plasma formed in ultrarelativistic heavy-ion collisions \cite{Heinz:2013th,Romatschke:2017ejr}. 
 From the perspective of modern effective field theory \cite{Weinberg:1978kz,Polchinski:1992ed}, hydrodynamics is considered an effective theory describing the long-wavelength, low-momentum dynamics of interacting quantum/classical many-body systems \cite{Kovtun:2012rj,Rocha:2023ilf}.
 Nonetheless, surprisingly enough, typical signatures commonly associated with hydrodynamic behavior (encoded in the anisotropic flow of hadrons \cite{Luzum:2013yya}) are found in systems as small as those formed in proton-proton and proton-nucleus collisions. This is remarkable given that in such systems there is no clear way to distinguish between micro and macroscopic length scales, as the inverse of the radius of the proton is comparable to the system's temperature \cite{Noronha:2024dtq} (in natural units $\hbar=c=k_B=1$).

The apparent hydrodynamic signatures displayed by exotic quantum systems, such as the quark-gluon plasma, seem to challenge the general belief \cite{Landau1987Fluid} that hydrodynamic behavior requires a very large number of degrees of freedom to meaningfully characterize the dynamics of conserved quantities in terms of local, \emph{average} fluid-like quantities such as temperature, flow velocity, and chemical potential. Given the recent experimental advances concerning the emergent hydrodynamic behavior of quantum systems with few degrees of freedom, such as quark-gluon droplets \cite{PHENIX:2018lia} and polariton fluids \cite{Su_rez_Forero_2020}, it is natural to wonder how many degrees of freedom are necessary for a quantum system to exhibit hydrodynamic behavior. Further support for the idea that quantum phenomena may lead to a broader regime of applicability for hydrodynamics can be found in important breakthroughs on the quantum origins of thermal behavior
\cite{PhysRevE.50.888,popescu2006entanglement,PhysRevLett.96.050403,rigol2008thermalization,kaufman2016quantum,Deutsch_2018,Mori_2018,doi:10.1126/science.aah5776,PhysRevResearch.2.043087}. 
In particular, recent work signals the emergence of thermal features in isolated quantum systems, even for a relatively small number of degrees of freedom 
\cite{kaufman2016quantum,doi:10.1126/science.aah5776,PhysRevResearch.2.043087}. 

Consider the case of a single, isolated quantum particle of mass $m$. The Schr\"odinger equation for the wave function $\psi(\vec{x},t)$ leads to a local form of probability conservation, described in terms of $\partial_t n + \nabla \cdot \vec{j} = 0$, where $n = |\psi|^2$ is the probability density and  $\vec{j} = -i\hbar(\psi^* \nabla \psi - \psi \nabla \psi^*)/m$ is the probability flux \cite{Sakurai_Napolitano_2020}. 
In fact, it has been known since Madelung \cite{Madelung:1927ksh} that one may \emph{exactly} rewrite the Schrödinger equation as a set of nonlinear ideal-fluid-like equations involving $|\psi|^2$ and a flow velocity defined by the gradient of the phase of the wave function (the flow is then irrotational). This Madelung representation revealed that the dynamics of the single particle wavefunction could be mathematically described through a hydrodynamic lens, with a quantum potential term encapsulating non-classical effects such as wave-particle duality and interference \cite{PhysRev.85.166}. The general understanding is that, through the Madelung transformation, the dynamical evolution of a quantum state can be interpreted as a geometric flow in the space of probability densities \cite{Khesin_2019}. Therefore, though insightful \cite{wyatt2005quantum}, the Madelung transform per se does not explain how hydrodynamics emerges as a universal, late-time, long-wavelength limit of quantum many-body systems. Furthermore, Madelung's result assumes isolated systems, and the effects of dissipation and environmental interactions remain an active field of research. For instance, previous works considered modifications of the Schr\"odinger equation to allow for dissipative effects (see, e.g., \cite{Sato:2022tbi,salasnich2024quantum}).

In this paper, we provide an explicit example of how decoherence in open quantum systems can lead to hydrodynamic behavior in the late-time, long-wavelength regime. We focus on the case of a single non-relativistic quantum particle linearly coupled to a bath of harmonic oscillators at temperature $T$, as described by the Caldeira-Leggett master equation \cite{Caldeira:1982iu}. Because of position-space decoherence, the reduced density matrix of the system becomes nearly diagonal at late times. By performing a series expansion of the reduced density matrix in powers of the off-diagonal spatial components, we obtain an infinite set of equations for the moments of the reduced density matrix that is analogous to the BBGKY hierarchy \cite{yvon1935theorie, bogoliubov1946kinetic, born1946general, 10.1063/1.1724117}. Truncating these moment equations at second order leads to a set of dissipative transient hydrodynamic equations, similar to those \cite{Denicol:2012cn} widely used in the study of the quark-gluon plasma \cite{Rocha:2023ilf}. These transient hydrodynamic equations describe how shear and bulk viscous flow emerge from decoherence in this open quantum system with transport coefficients directly determined by the damping constant $\gamma$, which encodes the influence of the environment. The asymptotic limit of these equations reduces to the celebrated Navier-Stokes equations \cite{Landau1987Fluid}, shown here to emerge from decoherence. Because the system is open, energy and momentum are not exactly conserved, which is manifested by the presence of new source terms in the hydrodynamic equations that are not usually considered in relativity, at least in the context of heavy-ion collisions  \cite{Rocha:2023ilf}. 

By linearizing the hydrodynamic equations around equilibrium, we compute the dispersion relations $\omega = \omega(k)$ of the corresponding modes (with $k$ being the wavenumber). A purely diffusive hydrodynamic ($\omega (k\to 0)=0$, i.e. ``gapless") mode $\omega \sim -i D k^2$, with a diffusion constant given by $D = k_B T/2m\gamma$ associated with the conservation law of probability, is found. The coupling to the environment leads to new stable non-hydrodynamic modes, with a gap $\omega (k \to 0) \sim -i \gamma$ \cite{PhysRevE.97.012130, PhysRevD.103.056020,10.1093/ptep/ptaa005}, in contrast to the behavior observed in Hamiltonian systems. Furthermore, using the Wigner function, the same hydrodynamic equations may be derived from a truncation of an effective kinetic theory using the method of moments \cite{grad:1949kinetic}. To further investigate this effective kinetic theory description and the onset of hydrodynamic behavior, we analytically solve the evolution equation of the Wigner function for the Caldeira-Leggett model and use it to calculate the analytical expressions for the hydrodynamic quantities, such as the velocity $\vec{u}$ and the viscous stress $\pi_{ij}$. Using these results, we show that the Navier-Stokes constitutive relations become a good description of the exact expressions for the dissipative fluxes at late times. This can be used to define the onset of hydrodynamic behavior in this open quantum system.

This paper is organized as follows. For the sake of completeness, in Section \ref{sec:Madelung equation}, we review the Madelung transform, which recasts the Schrödinger equation of an isolated quantum system in an ideal hydrodynamic form. In Section \ref{sec: Caldeira-Leggett Master Equation and Quantum Decoherence}, we briefly introduce the dynamics of open quantum systems and emphasize the role of quantum decoherence in the context of the Caldeira-Leggett model. In Section \ref{sec: Emergent Viscous Hydrodynamics}, we derive the viscous hydrodynamic equations and their transport coefficients for an open quantum system described by the Caldeira-Leggett master equation using our position-space decoherence series. In this section, we also present this system's hydrodynamic and non-hydrodynamic collective modes. In Section \ref{sec: Derivation from Kinetic Theory}, we establish a mapping between our expansion scheme and the cumulant expansion of the Wigner function, demonstrating how the same hydrodynamic equations can be derived from a truncation of an effective kinetic theory. We also investigate the onset of hydrodynamics as determined using an analytical solution of the Wigner function for the Caldeira-Leggett system. We present our conclusions and outlook in Section \ref{Conclusion}. In Appendix \ref{appendix: Caldeira-Leggett Model from Schwinger-Keldysh Formalism}, we derive the effective action of the Caldeira-Leggett model using the Schwinger-Keldysh formalism and demonstrate that our position-space decoherence expansion scheme effectively corresponds to the scenario where quantum fluctuations are small.
Finally, in Appendix \ref{appendix:Moments of Wigner Function}, we show in detail how to obtain the corresponding hydrodynamic equations by taking moments of the Wigner distribution function.

\section{Madelung Equation} \label{sec:Madelung equation}

In 1926, Madelung reinterpreted Schrödinger's equation for a single quantum particle into a more classical and easily visualizable form, drawing parallels with hydrodynamics \cite{Madelung:1927ksh}. The Madelung equations can be derived by expressing the wavefunction in polar form as follows
\begin{equation}
    \label{Eq:MadelungAnsatz}
    \psi(\vec{x},t) = \sqrt{n (\vec{x},t)}  e^{iS(\vec{x},t)/\hbar}, 
\end{equation}
where \(\sqrt{n}  \geq 0\) and \(S\) are both real-valued fields. The probability density is then given by $n = |\psi|^2$. Using this polar form, the probability flux becomes $\vec{J} =n\vec{u}$, where the flow velocity is defined as
\begin{equation}
    \vec{u} = \frac{1}{m} \nabla S.
\end{equation}
This velocity is not the particle's velocity but rather an effective field that characterizes how the probability density evolves in space and time.\footnote{This can be understood by analogy using the Huygens–Fresnel principle: (i.) every point on a wavefront acts as a source of secondary spherical wavelets, (ii.) the new wavefront at a later time is the envelope of all these secondary wavelets. The gradient of the phase is perpendicular to the wavefront and points in the direction of wave propagation, so that it can represent the velocity of the wave. Indeed, for a free particle we have $\vec{u} =\frac{\nabla S}{m} =\frac{\hbar }{m} \nabla (\vec{k}\cdot \vec{x} -\omega t)=\frac{\hbar\vec{k} }{m}=\frac{\vec{p} }{m}$.} Now, by substituting the polar form into the time-dependent Schrödinger equation 
\begin{equation}
    i\hbar \frac{\partial \psi(\vec{x},t)}{\partial t} = \left[ -\frac{\hbar^2}{2m} \nabla^2 + V(\vec{x}) \right] \psi(\vec{x},t),
\end{equation}
and after differentiating and separating the real and imaginary components, the following coupled partial differential equations arise 
\begin{equation}
    \frac{\partial \sqrt{n}}{\partial t} + \frac{1}{m} \nabla \sqrt{n} \cdot \nabla S + \frac{1}{2m} \sqrt{n} \nabla^2 S=0 ,
    \label{eq:madelung continuity equation}
\end{equation}
and
\begin{equation}
\frac{\partial S}{\partial t} + \frac{1}{2m} (\nabla S)^2 + V(\vec{x}) = \frac{\hbar^2}{2m} \frac{\nabla^2 \sqrt{n}}{\sqrt{n}}.
\label{eq:quantum Euler equation}
\end{equation}
The first equation corresponds to the imaginary part of the Schrödinger equation, which can be interpreted as the continuity equation for probability flow. The second equation, corresponding to the real part, is commonly referred to as the quantum Hamilton-Jacobi equation \cite{wyatt2005quantum}. To arrive at the Madelung equations, multiply the first equation by \(2\sqrt{n} \), and calculate the gradient of the second equation. This results in
\begin{equation}
    \frac{\partial n}{\partial t} + \nabla \cdot (n \vec{u}) = 0,
\end{equation}
and 
\begin{equation} \label{Eq:quantum_Euler}
     \frac{\partial \vec{u}}{\partial t} + (\vec{u} \cdot \nabla) \vec{u} = -\frac{1}{m} \nabla \left[V(\vec{x}) - Q(\vec{x},t) \right],
\end{equation}
where the term 
\begin{equation}
    Q(\vec{x},t) = -\frac{\hbar^2}{2m} \frac{\nabla^2 \sqrt{n(\vec{x},t)}}{\sqrt{n(\vec{x},t)}},
\end{equation}
is recognized as the Bohm quantum potential \cite{PhysRev.85.166}. In the context of this hydrodynamic formulation, the quantum potential governs the motion of the system without direct reference to classical forces, being present only when $\hbar \neq 0$. Unlike classical potentials, the quantum potential depends on the global structure of the wave function, allowing for the description of well-known quantum phenomena, such as quantum tunneling.\footnote{Consider a particle approaching a potential barrier $V(x)$ of height $V_0$. In the Madelung view, the quantum potential $Q$ modifies the effective potential energy landscape experienced by the particle, where $V_{\text{eff}}=V+Q$. The particle moves according to $\frac{\partial S}{\partial t} + \frac{1}{2m} \left( \frac{\partial S}{\partial x} \right)^2 + V_{\text{eff}}(x) = 0$. Even if $E < V_0$, the quantum potential $Q(x)$ can make $V_{\text{eff}}(x)$ lower than $V_0$ in certain regions, allowing the particle to traverse the barrier. Therefore, one can view quantum tunneling as a consequence of the quantum potential modifying the effective potential energy landscape.}

One can study the emergent collective excitations following from the Madelung equations by considering small perturbations around a stationary state
\begin{equation}
    n(\vec{x},t) = n_0(\vec{x}) + \delta n(\vec{x},t), \quad \vec{u}(\vec{x},t) = \vec{0} + \delta \vec{u}(\vec{x},t).
\end{equation}
Notably, unlike classical fluids, the equilibrium probability density of this quantum fluid cannot be a constant for all space, as this would render the corresponding wave function non-normalizable. The stationary state one may consider could be as simple as that of a particle confined within a potential well, as long as we are interested in disturbances far from the edges (where, for simplicity, $V$ may be assumed to vanish). Following these assumptions, we find the following fourth-order partial differential equation for $\delta n(\vec{x},t)$
\begin{equation}
\label{eq:Madelung2ndOrder}
    \partial_t^2 \delta n + \frac{\hbar^2}{4m^2} \nabla \cdot \left[ n_0 \nabla \left( \frac{1}{n_0} \nabla^2 \delta n - \frac{\nabla^2 n_0}{n_0^2} \delta n - \frac{2 \nabla n_0}{n_0^2} \cdot \nabla \delta n \right) \right] = 0.
\end{equation}
Assuming a time–harmonic dependence of the fluctuations
\begin{equation}
    \delta n(\vec{x}, t) = \delta n_n(\vec{x}) e^{-i \omega t},
\end{equation}
we obtain the eigenvalue equation
\begin{equation}
\nabla \cdot \left\{ n_0 \nabla \left[ 
\frac{1}{n_0} \nabla^2 \delta n_n
- \frac{\nabla^2 n_0}{n_0^2} \delta n_n 
- 2 \frac{\nabla n_0}{n_0^2} \cdot \nabla \delta n_n
\right] \right\} = \frac{4 m^2 \omega^2}{\hbar^2} \delta n_n.
\end{equation}
As mentioned above, $n_0$ is not constant, but we will assume it varies slowly on the length scales of the perturbations (so that $\nabla n_0$ terms can be neglected in this approximation). Then, the above eigenvalue equation gives the following dispersion relation 
\begin{equation}
    \omega(k) =\pm \frac{\hbar k^2}{2m}.
    \label{eq:Madelung modes}
\end{equation}
This describes stable, oscillatory behavior with strong dispersion, where phase and group velocities increase linearly with $k$. One may wonder why there are two modes, while we would only obtain a single mode from the Schrödinger equation. The root of this confusion lies in the \textit{time-reversal symmetry} of the Schrödinger equation. Eq.~\eqref{eq:Madelung2ndOrder} is obviously invariant under time reversal $t\to -t$, which means that for each solution there is a corresponding time-reversed solution with $\omega\to-\omega$. In the original Schr\"odinger equation, on the other hand, time reversal is only a symmetry if combined with a complex conjugation operation, $\psi \to \psi^*$, so that $i\omega \to i\omega$ is left invariant. 

Finally, the circulation of the flow velocity field along any closed path obeys the auxiliary quantization condition 
\begin{equation}
    \Gamma \mathrel= \oint m \vec{u} \cdot d\vec{l} = 2\pi n \hbar,
    \label{eq:quantization}
\end{equation}
where $n\in \mathbb{Z}$. Note that the velocity field is generally irrotational because it is derived from the gradient of the scalar phase $S$, implying $\nabla \times \vec{u} =\nabla \times \left ( \nabla S/m \right ) =0$. Different behavior emerges in the presence of quantum vortices where $S$ is singular at $n=0$, which means the circulation $\Gamma \ne 0$. To prove Eq.~\eqref{eq:quantization}, recall the definition of the fluid velocity and write $\vec{u}=\frac{\nabla S}{m} =\hbar\frac{ \nabla\phi }{m}$. This leads to $\Gamma = \hbar \oint \nabla \phi \cdot d\vec{l} = \hbar \Delta \phi$. Since the wavefunction must be single-valued, the change in phase $\Delta\phi$ around a closed loop must be an integer multiple of $2\pi$: $\Delta\phi=2\pi n$, $n\in \mathbb{Z}$. Thus, the circulation is quantized $\Gamma \mathrel=2\pi n \hbar$.\footnote{For readers interested in learning more about the mathematical work on the Madelung equations, see \cite{Reddiger:2022fsd, Khesin_2019, Foskett_2024, fusca2016madelungtransformmomentummap}.}

\section{Caldeira-Leggett Master Equation and Quantum Decoherence} \label{sec: Caldeira-Leggett Master Equation and Quantum Decoherence}
\subsection{The dynamics of open quantum systems}

While revealing the proximity between the quantum mechanics of a single particle and ideal hydrodynamics, the above discussion is limited to a zero-entropy pure quantum state of an isolated quantum particle undergoing unitary evolution.  
A more general and realistic account of the evolution of a particle should include the effects of interactions with its surroundings. 
Hence, entanglement with degrees of freedom of an uncontrolled environment would lead to loss of information and thus non-unitary quantum dynamics \cite{Rivas_2012}. Investigating the emergence of hydrodynamic behavior in such an open quantum system is the main focus of this paper.  

The description of an open quantum system requires that we consider the possibility of mixed states, generically represented by a density operator $\rho$ such that the expectation value of an observable $\mathcal{O}$ is given by $\langle \mathcal{O}\rangle=\tr \left ( \rho \mathcal{O}  \right )$.  
We start from a density operator $\rho_{\text{total}}$ describing both the system $S$ of interest and its environment $E$. 
Since the total system composed of $E+S$ is closed, its evolution is unitary and given by the von Neumann equation \cite{10.1093/acprof:oso/9780199213900.001.0001}
\begin{equation}
    \frac{d }{dt}\rho_{\text{total}}\left ( t \right )  = -\frac{i}{\hbar} [H_{\text{total}}\left ( t \right ) , \rho_{\text{total}} \left ( t \right ) ],
    \label{eq: Von-Neumann}
\end{equation}
where $H_{\text{total}}\left ( t \right )$ is the complete Hamiltonian, 
\begin{equation}
    H_{\text{total} } =H_S+H_E+H_{\text{int} },
\end{equation}
with $H_S$ being the Hamiltonian of the system of interest, $H_E$ is the Hamiltonian of its environment $E$, and $H_{\text{int}}$ describes the coupling between both of them. 

As we are interested only in the dynamics of the system, we trace out the environment's degrees of freedom to obtain the reduced density matrix
\begin{equation}
    \rho_S = \tr_E \left\{ \rho_{\text{total}} \right\},
\end{equation}
which is used to compute averages and correlations involving any observables of the system $S$ alone --- for instance, the expectation value $\langle \mathcal{O}\rangle=\tr \left ( \rho_S \mathcal{O}_S  \right )$ of an operator $\mathcal{O} = \mathcal{O}_S \otimes \mathds{1}$. 
Here, $\tr_E\left( \cdots \right) = \sum_i \langle e_i | \cdots | e_i \rangle.$ denotes the \textit{partial trace} over states of the environment, as given by a complete basis $\{ |e_i\rangle \}$ for its Hilbert space. The challenge now is to find an equation governing the evolution of the reduced density matrix $\rho_S$.

In this paper, we consider the case where $\rho_S$ is governed by the Caldeira–Leggett model \cite{Caldeira:1982iu}. For the sake of completeness, in the following we briefly review the features of this model that are relevant to this work. The Caldeira-Leggett model provides a concrete framework for studying quantum dissipation; it models the environment at finite temperature $T$ as an infinite set of harmonic oscillators linearly coupled to the system. The oscillators represent environmental degrees of freedom like phonons in a solid or modes of an electromagnetic field. Assuming the system is a single quantum particle in a potential $V(\vec{x})$, the total Hamiltonian is given by
\begin{equation}
\begin{split}
    H_{\text{total}} &=\frac{p^2}{2m} + V(\vec{x})+\sum_n \left( \frac{p_n^2}{2m_n} + \frac{1}{2}m_ n \omega_n^2 x_n^2 \right)- \vec{x} \cdot \sum_n c_n \vec{x}_n + x^2 \sum_n \frac{c_n^2}{2m_n \omega_n^2}\\
    &=H_S + H_E + H_{\text{int}}+H_C,
\end{split}
\end{equation}
where $c_n$ is the coupling constant between the system and the $n$-th oscillator. The interaction Hamiltonian $H_{\text{int}}$ describes a continuous monitoring of the position of the system by the environment. The appearance of the final counter-term, $H_C \sim x^2$, is due to the interaction between the system and high-frequency modes in the environment, which leads to a renormalization of the system's potential. The counter-term is introduced to prevent unphysical divergences, in particular, by ensuring that the system's energy remains finite and well-behaved via corrections to the bare potential $V(\vec{x})$. This is analogous to calculations in quantum field theory \cite{Coleman:2018mew}, where formally infinite contributions from interactions are systematically absorbed into physical parameters.  

To derive the master equation for the reduced density matrix $\rho_S$ of the system, one employs the Feynman-Vernon influence functional method \cite{FEYNMAN1963118} to integrate out the bath degrees of freedom, and models the environment with an Ohmic spectral density with a Lorentz-Drude cutoff function
\begin{equation}
    J(\omega) = \frac{2m\gamma}{\pi} \omega \frac{\Omega^{2}}{\Omega^{2} + \omega^2},
\end{equation}
where $\gamma$ is the damping constant and $\Omega$ is a high-frequency cutoff. The spectral function, $J(\omega)$, describes how the system responds to spontaneous thermal fluctuations. To derive the equation of motion for $\rho_S$, we assume the high-temperature limit $k_BT\gg \hbar \omega$ and the weak coupling assumption $\hbar \gamma \ll \text{Min}   \left \{ \hbar \Omega, 2\pi k_BT  \right \}$. Furthermore, we focus on the case where the characteristic timescale of the system is slow in comparison to the bath correlation time $\hbar \omega _0 \ll \text{Min}   \left \{ \hbar \Omega, 2\pi k_BT  \right \} $, where $\omega_0$ is the typical frequency of the system's evolution. In summary, the separation of the timescales satisfy $\tau_\pi\gg \tau_B$ and $\tau_S \gg \tau_B$ where $\tau_B\equiv \text{Max} \left \{ \Omega^{-1},\hbar /2\pi k_BT \right \}$ is the correlation time, $\tau_\pi \equiv1/4\gamma$ is the relaxation time, and $\tau_S \equiv \omega_0^{-1}$ is the typical timescale for the system. Under these conditions, the influence functional simplifies, allowing us to obtain the celebrated Caldeira-Leggett master equation  \cite{Caldeira:1982iu}
\begin{equation}
    \frac{\partial }{\partial t}\rho_S\left ( t \right )  = -\frac{i}{\hbar}[H_S, \rho_S\left ( t \right )] - \frac{i \gamma}{\hbar}[x, \{p, \rho_S\left ( t \right )\}] - \frac{2m\gamma k_B T}{\hbar^2} [x, [x, \rho_S\left ( t \right )]].
    \label{eq: Caldeira-Leggett}
\end{equation}
This is the dynamical equation for the open quantum system, describing the system's decoherence and dissipative behavior due to its interaction with the environment. The first term on the right-hand side of the master equation is the coherent, unitary evolution of the system, which encapsulates how it would evolve on its own, without the influence of decoherence or dissipation from the environment. The second term introduces dissipation into the system, playing the role of a frictional force, representing the energy loss from the system to the environment. The last term is essential because it directly accounts for quantum decoherence \cite{zeh1970interpretation}. Decoherence refers to the process by which quantum superpositions (off-diagonal elements of the density matrix) decay due to interacting with an external environment. 

The Caldeira-Leggett master equation in the coordinate representation is given by
\begin{align}
\begin{split}
\label{eq:Caldeira-Legget-x}
\frac{\partial}{\partial t} \rho_S(\vec{x}, \vec{x}\,', t) &=  \frac{i \hbar}{2m} \left( \nabla^2 - \nabla\,'^2 \right) \rho_S(\vec{x}, \vec{x}\,', t) - \frac{i}{\hbar} (V(\vec{x}) - V(\vec{x}\,'))  \rho_S(\vec{x}, \vec{x}\,', t) \\
&\quad - \gamma (\vec{x} - \vec{x}\,') \left( \nabla - \nabla\,' \right) \rho_S(\vec{x}, \vec{x}\,', t) - \frac{2m \gamma k_B T}{\hbar^2} (\vec{x} - \vec{x}\,')^2 \rho_S(\vec{x}, \vec{x}\,', t),
\end{split}
\end{align}
where $\rho_S( \vec{x}, \vec{x}\,',t)=\langle\vec{x} | \rho _S\left ( t \right )  |\vec{x}\,'   \rangle $ is the density matrix in coordinate representation. Again, the third term on the right-hand side is called the damping term, modeling the dissipative interaction with the environment, affecting the coherence between positions $\vec{x}$ and $\vec{x}\,'$. The fourth term is a diffusion term, representing noise or thermal fluctuations from the environment, causing the quantum state to spread and decohere, particularly affecting the off-diagonal elements in position space. 

Unlike the first two terms, the last two terms in Eq.~\eqref{eq:Caldeira-Legget-x} are not of the form $\big(F(\vec x, \nabla) - F(\vec x\,', \nabla')\big) \rho_S(\vec{x}, \vec{x}\,', t)$. Therefore, as a result of the partial trace, Eq.~\eqref{eq:Caldeira-Legget-x} admits no solution of the form $\rho_S=\psi^*(\vec x\,',t)\psi(\vec x,t)$ for $\gamma\neq 0$ \cite{Caldeira:1982iu}. While this prevents a straightforward generalization of the Madelung equations, we will see that hydrodynamic equations still apply at late times, as off-diagonal contributions to $\rho_S(\vec{x}, \vec{x}\,', t)$ are suppressed.

\subsection{The evolution of off-diagonal elements and decoherence}

To investigate the evolution of off-diagonal elements over time, we turn to the concept of decoherence in quantum mechanics (for a review about decoherence, see, e.g. \cite{Schlosshauer_2019}). In its most fundamental form, decoherence emerges from specific types of system-environment interactions. A key characteristic of such interactions is that the reduced system influences the environment in a way that establishes system-environment correlations. However, the feedback from the environment to certain system states remains negligibly small. Consequently, the population damping of the reduced density matrix in a particular basis is minimal, while the coherences typically decay rapidly over very short timescales.

In the Caldeira-Leggett model, environmental interactions cause the reduced density matrix $\rho_S(\vec{x}, \vec{x}\,', t)$ to evolve such that the off-diagonal elements where $\vec{x} \neq \vec{x}\,'$, representing quantum coherences between different positions, decay exponentially due to decoherence.
This decay occurs because the environment effectively acts as a measuring device, collapsing superposition states and driving the system toward classical behavior, so the decoherence can thus be thought of as a process arising from the continuous monitoring of the system by the environment \cite{PhysRevD.24.1516}. 

For sufficiently high temperatures or large coherent separations $\Delta x=\left | \vec{x}-\vec{x}\,'  \right |$, the final term on the right-hand side of Eq.~\eqref{eq: Caldeira-Leggett} dominates. 
In terms of the thermal de Broglie wavelength $\lambda_{\text{th}} = \hbar/\sqrt{2 m k_B T}$, 
Eq.~\eqref{eq:Caldeira-Legget-x}  becomes
\begin{equation}
    \frac{\partial}{\partial t} \rho_S(\vec{x}, \vec{x}\,', t) \sim  -  \gamma \left( \frac{\vec{x}- \vec{x}\,'}{\lambda _{\text{th} }} \right)^2 \rho_S( \vec{x}, \vec{x}\,',t)\,,
    \quad \text{ for }
    |\vec x -\vec x\,'|\gg \lambda_{\text{th}}
    \label{Eq:xspaceMasterEq}
\end{equation}
Thus, elements of the density matrix at large distances from the diagonal (in comparison to $\lambda_{\text{th}}$) decay exponentially according to 
\begin{equation}
   \rho_S(\vec{x}, \vec{x}\,', t) \sim  e^{\Gamma \left ( t \right ) } \rho_S(\vec{x}, \vec{x}\,', 0)\,,
   \quad \text{ for }
   |\vec x -\vec x\,'|\gg \lambda_{\text{th}}\,
    \label{eq:decoherence}
\end{equation}
where $\Gamma \left ( t \right ) =- \Lambda \left | \vec{x} -\vec{x}\,' \right | ^2t$ is the decoherence function and $\Lambda \equiv \gamma/\lambda _{\text{th}}^2$ plays the role of a decoherence rate. 
This exponential decay is only expected for coherent separations $\Delta x$, much larger than $\lambda_{\text{th}}$. 
For  $\Delta x \lesssim \lambda_{\text{th}}$, on the other hand, we expect coherences in the position representation  to remain unsuppressed. 
In fact, elements of $\rho_S$ can be shown to be distributed according to a Gaussian of width $\sqrt{2} \lambda_{\text{th}}$ around the diagonal in equilibrium \cite{10.1093/acprof:oso/9780199213900.001.0001}.

The dependence of the characteristic decoherence time $\tau_D= \gamma^{-1} \lambda_{\text{th}}^2/ \Delta x^2$ on the coherent separation $\Delta x$ is reasonable \cite{Schlosshauer_2019}, given that $\Delta x$ represents the separation between different states of the quantum system. If the typical wavelengths of the  environment are much larger than $\Delta x$, the environment cannot easily distinguish between these separated states. In such a case, many interactions with the environment are needed for it to gather enough information to discern the separation between the quantum states, leading to slow decoherence. When $\Delta x$ grows larger, the environment can more readily distinguish between the different states of the quantum system. Thus, the information about the paths becomes more apparent with each scattering event, and the rate of decoherence increases. More interactions lead to faster loss of quantum coherence. Finally, if $\Delta x$ becomes larger than the typical environmental wavelength, the environment can easily distinguish the different states, even in a single scattering event. In this situation, the decoherence rate no longer depends on $\Delta x$, as the environment can distinguish the states with maximum efficiency. This is the so-called ``short-wavelength limit,'' where decoherence occurs as quickly as possible.

\begin{figure}
    \centering
    \includegraphics[width=0.9\linewidth]{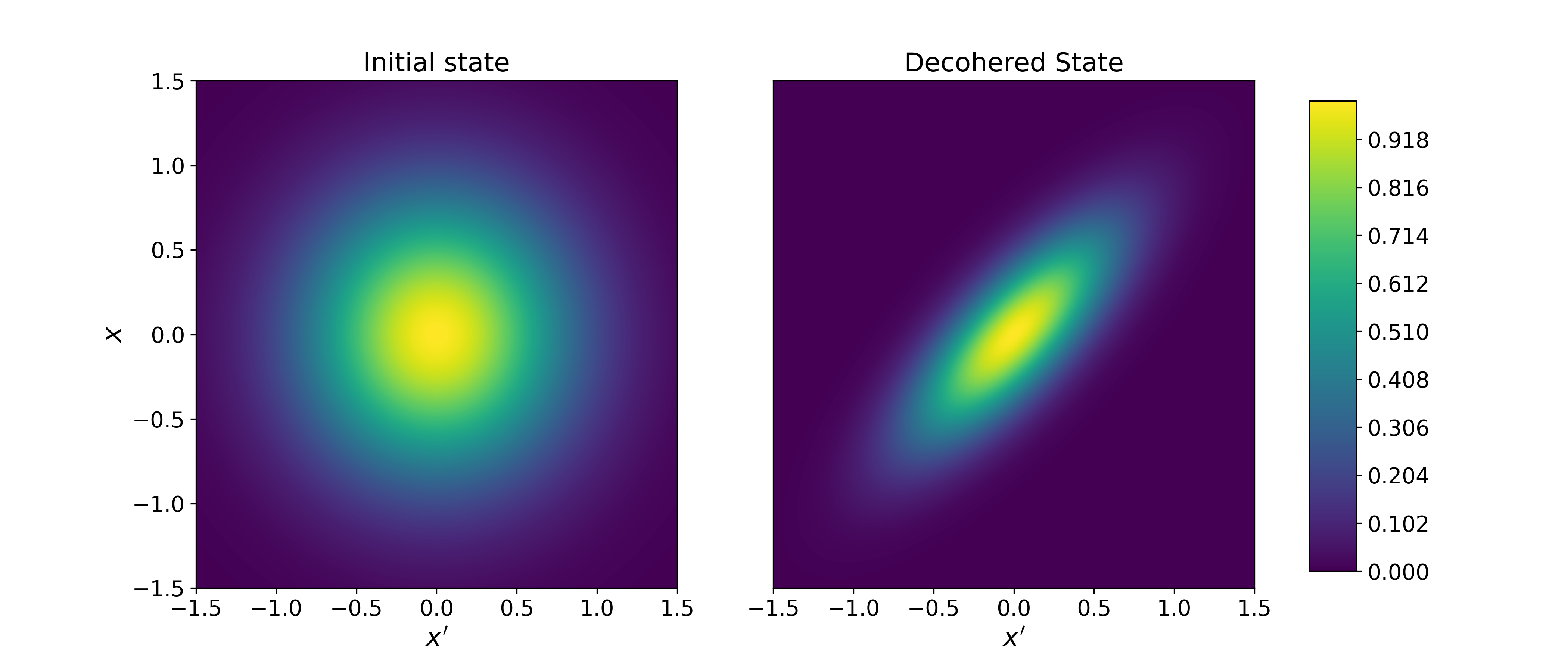}
    \caption{Illustration of decoherence for a density matrix representing a Gaussian wave packet. As one moves from left to right, the spatial coherence length—indicated by the Gaussian's width along the off-diagonal axis $x=-x\,'$ gradually diminishes due to environmental interactions.
    }
    \label{fig:decoherence}
\end{figure}

To give some intuition about the effects from decoherence, let us discuss the following case. Consider a free particle in one spatial dimension, ignoring the dissipation term in the Caldeira-Leggett master equation. 
For the sake of illustration, we begin with a Gaussian wave packet centered at $x=0$. The evolution over time is depicted in Fig.~\ref{fig:decoherence}. As time progresses, the coherence length, defined as the Gaussian's width along the off-diagonal axis $x\,'=-x$, which represents spatial coherence, diminishes. This illustrates the process of decoherence. Consequently, the density matrix transitions toward a quasi-classical probability distribution concentrated along $x=x\,'$. Meanwhile, the spread of the ensemble along the diagonal direction $x=x\,'$, which indicates the size of the probability distribution $\rho_S\left ( x,x\,'=x,t \right )$, increases over time.
In our previous approximate solution, because of the neglect of damping effects, the diagonal elements of the density matrix were not affected in the course of the time evolution, whereas the off-diagonal elements quickly decayed. Now note that according to Eq.~\eqref{eq:decoherence}, the elements farther from the diagonal decay faster, while elements with $\Delta x \lesssim \lambda_{\text{th}}$ are not necessarily suppressed. Therefore, we conclude that when $t\gg \tau _D$, the reduced density matrix will be concentrated around the diagonal, within a typical distance given by the thermal de Broglie wavelength $\lambda_{\text{th}}$.

This kind of behavior is just an approximation, as the behavior of the decoherence function $\Gamma \left ( t \right ) =-\Lambda \Delta x^2t$ depends on many factors \cite{10.1093/acprof:oso/9780199213900.001.0001}. Generally speaking, the strength of the decoherence effect depends both on its initial quantum state and on the Hamiltonian that governs the interaction between the system and environment. However, one ultimately expects exponential decay of the off-diagonal elements. A more rigorous result can be obtained in the high temperature limit \cite{10.1093/acprof:oso/9780199213900.001.0001}
\begin{align}
    \Gamma \left ( t \right ) \approx -\frac{1}{3} \frac{\Lambda }{m^2} \Delta p^2t^3,
\end{align}
where $\Delta p$ is the initial separation of the superposed wave packets in momentum space. Thus, one can  assert that the off-diagonal elements of the reduced density matrix decay exponentially to zero within an extremely short timescale. 
This rapid decay, driven by decoherence, provides a \textit{physical explanation} for our approach of performing a Taylor expansion around the diagonal elements of the system's density matrix in the next section.

\section{Emergent Viscous Hydrodynamics of the Caldeira-Leggett Model} \label{sec: Emergent Viscous Hydrodynamics}

We now derive the hydrodynamic limit encoded in the Caldeira-Leggett master equation, which will govern the late-time, long-wavelength dynamics of a quantum particle interacting with its environment. Inspired by Madelung's hydrodynamic formulation of quantum mechanics, we begin by adopting a similar approach for the system's density matrix. Specifically, we express the density matrix in a polar form as follows
\begin{align}
    \rho_S =e^{R+iS},
    \label{eq:polar}
\end{align}
where $R, S : (\vec{x}, \vec{x}\,', t) \in \mathbb{R}^3 \times \mathbb{R}^3 \times \mathbb{R}^+ \longmapsto \mathbb{R}$ are real-valued functions representing the logarithm of the amplitude and the phase of the density matrix, respectively. Because the density matrix is Hermitian, $\rho_S(\vec x, \vec x\,')=\rho_S(\vec x\,', \vec x)^*$, it follows that $R$ is symmetric, while $S$ is antisymmetric under $\vec x \leftrightarrow \vec x\,'$.

Substituting Eq.~\eqref{eq:polar} into the Caldeira-Leggett master equation and performing algebraic manipulations, we obtain the following coupled equations for $R$ and $S$
\begin{equation}
     \partial _t R=-\frac{\hbar }{2m} \left ( \nabla _+\cdot \nabla _-S+\nabla _+ R\cdot \nabla _-S+\nabla _+S\cdot \nabla _-R \right )-2\gamma \vec{x}_{-}\cdot \nabla _-R-\frac{8m\gamma k_BT}{\hbar ^2} \vec{x}_{-}^{\,2},
     \label{eq:real part}
\end{equation}
\begin{equation}
        \partial _tS=\frac{\hbar }{2m}   \left ( \nabla _+\cdot \nabla _-R+\nabla _+R\cdot \nabla_-R-\nabla _+S\cdot \nabla _-S  \right )-\frac{1}{\hbar }\left(V( \vec{x}_+ + \vec{x}_-) - V( \vec{x}_+ - \vec{x}_-)\right)  -2\gamma\, \vec{x}_-\cdot \nabla _-S,
        \label{eq:imaginary part}
\end{equation}
where we have introduced the center-of-mass and relative coordinates 
\begin{equation} 
\vec{x}_+ = \frac{\vec{x} + \vec{x}\,'}{2}, \quad \vec{x}_- = \frac{\vec{x} - \vec{x}\,'}{2}, 
\end{equation}
and the differential operators
\begin{equation}
    \nabla _{-}=\nabla -\nabla\,', \quad \nabla _{+}=\nabla +\nabla\,'.
\end{equation}
To make these equations more tractable and to reveal the emergent hydrodynamic behavior, we employ a well-motivated expansion suitable for open quantum systems experiencing position-space induced decoherence.

\subsection{New series expansion for position-space decoherence}\label{Sec:IV.A}

We already know that due to the interaction between the system and the environment, far-off-diagonal elements of the density matrix decay exponentially. When the timescale $\tau_S$ we are considering is much larger than the decoherence timescale $\tau_D$, the system's density matrix approaches a nearly diagonal form. Therefore, decoherence effectively projects the system onto a low‐resolution, near‐diagonal subspace of the density matrix. This motivates us to consider performing a Taylor expansion of the density matrix around its diagonal elements. By expanding about $\vec{x}' = \vec{x}$, we write the density matrix as
\begin{equation}
\rho_S(\vec{x}, \vec{x}\,', t) = \rho_S(\vec{x}, \vec{x}, t) + \sum_{i} \left. \frac{\partial \rho_S}{\partial x_i} \right|_{\vec{x}= \vec{x}\,'} (x_i - x_i') + \frac{1}{2} \sum_{i,j} \left. \frac{\partial^2 \rho_S}{\partial x_i \partial x_j} \right|_{\vec{x}= \vec{x}\,'} (x_i - x_i')(x_j - x_j') + \cdots\,.
\end{equation}
Since we can write the density matrix in polar coordinate form, the real and imaginary parts are expanded to obtain
\begin{equation}
S=a_ix_{-}^i+B_{ijk}x_{-}^ix_{-}^jx_{-}^k+\cdot \cdot \cdot=S^{\left ( 1 \right )}+S^{\left ( 3 \right )}+\cdot \cdot \cdot ,
\label{eq:expansion of S}
\end{equation}
\begin{equation}
    R=\log n+C_{ij}x_{-}^ix_{-}^j+E_{ijkl}x_{-}^ix_{-}^jx_{-}^kx_{-}^l+\cdot \cdot \cdot =R^{\left ( 0 \right )}+R^{\left ( 2 \right )}+R^{\left ( 4 \right )}+\cdot \cdot \cdot ,
    \label{eq:expansion of R}
\end{equation}
where we employ the Einstein summation convention and note that $R$ is even and $S$ is odd under $\vec{x}_- \to - \vec{x}_-$. Above, 
because the probability density $n(\vec x)=\rho_S(\vec x,\vec x)$ is given by the density matrix with $\vec{x}_-=0$, we write the leading term in Eq.~\eqref{eq:expansion of R} as $\log n(\vec x_+)$. 
Note that the coefficients of the expansion, such as $a_i$, can only depend on the center-of-mass coordinate $\vec{x}_+$. This expansion is general and well-motivated in scenarios where decoherence in the position basis suppresses quantum coherence between spatially separated states. 
Because we expand $\log \rho_S$, at order $\mathcal{O}(|\vec x_-|^2)$ and above, our expansion can also capture the distribution of elements of $\rho_s$ for  $|\vec x_-|\lesssim \lambda_{\text{th}}$ close to the diagonal at long times, which approaches a Gaussian as equilibrium is approached.

The structure of Eqs.~\eqref{eq:real part} and \eqref{eq:imaginary part} always couples coefficients of different orders in $\vec{x}_-$, giving their dynamics a hierarchical structure reminiscent of the famous BBGKY hierarchy \cite{yvon1935theorie,bogoliubov1946kinetic, born1946general, 10.1063/1.1724117}. By systematically expanding in powers of $\vec{x}_-$ and truncating the series up to a given order,  we may capture the coarse-grained dynamics of the Caldeira-Leggett model in the regime where off-diagonal elements are negligible. 
From an effective field theory perspective, one can view this truncation as akin to integrating out short‐range quantum fluctuations, leaving an effective description in terms of slow, coarse‐grained variables. In fact, in Appendix \ref{appendix: Caldeira-Leggett Model from Schwinger-Keldysh Formalism} we discuss the interpretation of the variables $\vec{x}_+$ and $\vec{x}_-$ from the perspective of the Schwinger-Keldysh closed-time path formalism. 
Analogous to the gradient expansion in hydrodynamics \cite{Rocha:2023ilf}, each successive order in $\vec{x}_-$ encodes more refined corrections that become increasingly suppressed at large timescales. In the following sections, we delve into the physics associated with different orders in the expansion, exploring how they contribute to the emergent hydrodynamic behavior and how viscous terms arise naturally in this framework.

The truncation of the expansion in $\vec{x}_-$ is self‐consistent precisely because decoherence prevents the regrowth of long‐range coherences in the open system dynamics. In other words, higher‐order terms in $\vec{x}_-$ correspond to correlations that are rapidly washed out by the environment and, thus, become negligible on timescales much larger than $\tau_D$.

\subsection{Zeroth order term and local conservation of probability}
\label{sec:Expansion Zeroth order}

Let us first consider Eq.~\eqref{eq:real part}. To determine the zeroth order term $\mathcal{O}\left ( x_{-}^0 \right )$, we isolate the contributions proportional to $x_{-}^0$. We obtain
\begin{align}
    \partial _t\log n=-\frac{\hbar }{2m} \left ( \nabla _+\cdot \vec{a}+\vec{a}\cdot \nabla _+\log n   \right ).
\end{align}
One may rewrite this equation as a conservation law
\begin{align}
    \partial _t n+\nabla_+\cdot \left ( n\vec{u}  \right )=0,
    \label{eq:conservationlawdensity}
\end{align}
where the velocity is
\begin{align}
    \vec{u} =\frac{\hbar }{2m} \vec{a} .
    \label{eq:velocity}
\end{align}
Note that for any differentiable function, $\nabla_+ f(\vec{x}_+) \approx \nabla f(\vec{x})$ when $\vec{x} \to \vec{x}\,'$. The variable $\vec{x}_+$ describes the coarse-grained dynamics, while $\vec{x}_-$ captures quantum fluctuations. This implies that we are effectively dealing with a situation where the quantum fluctuations in the physical system are very weak and the system becomes effectively local. Therefore, if we only consider the zeroth order $\mathcal{O}\left ( x_{-}^0 \right )$, we obtain the continuity equation of probability, which shows that the probability is \textit{locally conserved}.

\subsection{First order term and the momentum equation}

Next, we consider terms of first order in $\mathcal{O}\left ( x_{-} \right )$. By isolating these terms in Eq.~\eqref{eq:imaginary part}, we find
\begin{equation}
    \partial _t\vec{u} +\left ( \vec{u}\cdot \nabla _+ \right ) \vec{u} =\frac{1}{mn} \nabla _{+}\cdot\mathcal{C}-\frac{1}{m}\nabla _{+}V-2\gamma \vec{u},
        \label{eq:momentum equation0}
\end{equation}
where we have also used Eq.~\eqref{eq:velocity}. 
We can write the above equation in a way that resembles the Euler equation
\begin{equation}
\partial_t (mn \vec{u}) + \nabla_+ \cdot \left(mn \vec{u} \otimes \vec{u} - \mathcal{C}\right) = -n \nabla_+ V - 2\gamma mn \vec{u},
    \label{eq:momentum equation}
\end{equation} 
where we have defined
\begin{equation}
    \mathcal{C}_{ij}\equiv \frac{n \hbar ^2}{2m} C_{ij}.
    \label{eq:quantum Cauchy stress tensor}
\end{equation}

Equation \eqref{eq:momentum equation} describes the dynamics of the momentum density of the open quantum system, including a damping force and viscous effects.  The striking resemblance of this equation to the classical Cauchy momentum equation justifies interpreting $\mathcal{C}_{ij}$ as the quantum analog of the Cauchy stress tensor. 
 That is, the comoving derivative of the momentum density $n\vec{u}$ equals the negative divergence of the stress tensor with additional terms that break momentum conservation, due to conservative and frictional forces. 
The appearance of the final term, $\sim \gamma \vec{u}$, signals the breaking of Galilean invariance due to the preferred velocity of the environment. This term stems from the coupling between the reduced system and the environment, which results in information leakage from the former to the latter. 
The violation of momentum conservation is similar to the case in electron hydrodynamics where the scattering of electrons off impurities and/or phonons leads to momentum non-conservation \cite{Lucas_2018}.

Note that the momentum equation Eq.~\eqref{eq:momentum equation0} in this form suggests that quantum effects such as those from the Bohm quantum potential should be contained within $\mathcal{C}$.  

\subsection{Higher order terms and truncation of hierarchy}

Finally, let us consider the second-order $\mathcal{O}\left ( x_{-}^2 \right )$ in Eq.~\eqref{eq:real part}. Isolating terms proportional to $x_{-}^ix_{-}^j$, one obtains the following equation
\begin{align}
    \partial _tC_{ij}=-\left ( \vec{u}\cdot \nabla _+  \right ) C_{ij}-\left ( \nabla _{+}^k u_i\right )C_{kj}-C_{ik }\left ( \nabla _{+}^k u_j\right )-\frac{\hbar }{2m}\nabla _{+}^kB_{ijk}-\frac{3\nabla _{+}^k n}{2mn}B_{ijk}-4\gamma C_{ij}-\frac{8m\gamma k_BT}{\hbar ^2}\delta  _{ij}.
\end{align}
This equation is evidently not closed due to the hierarchical structure of the full system of equations. In order to compute $C_{ij}$, we must know $B_{ijk}$, the equation of motion of which depends on tensors of higher rank, forming an infinite set of coupled equations. To obtain a closed set of equations, one may perform a simple \textit{truncation} motivated by the fact that $B_{ijk}$ comes from higher-order terms in the expansion parameter $|\vec x_-|$. Hence, we arrive at 
\begin{align}
    \partial _t C_{ij}=-\left ( \vec{u}\cdot \nabla _+  \right ) C_{ij}-\left ( \nabla _{+}^k u_i\right )C_{kj}-C_{ik }\left ( \nabla _{+}^k u_j\right )-4\gamma C_{ij}-\frac{8m\gamma k_BT}{\hbar^2}\delta _{ij},
    \label{eq:equation of motion for $C_{ij}$}
\end{align}
which has the form of a relaxation equation for  $C_{ij}$. 
From Eqs.~\eqref{eq:quantum Cauchy stress tensor} and \eqref{eq:equation of motion for $C_{ij}$} we can find the relaxation equation for the Cauchy stress tensor,
\begin{equation}
    \partial _t\mathcal{C}_{ij}+\left ( u_l\nabla _{+}^{l} \right ) \mathcal{C}_{ij}=-\left ( \nabla _{+}^k u_k\right )\mathcal{C}_{ij}-\left ( \nabla _{+}^k u_i\right )\mathcal{C}_{kj}-\mathcal{C}_{ik }\left ( \nabla _{+}^k u_j\right )-4\gamma \mathcal{C}_{ij}-4\gamma n k_BT\delta _{ij},
    \label{eq:stress tensor equation}
\end{equation}
which describes how $\mathcal{C}$ reaches its equilibrium value over time.

The above calculation indicates that, in quantum mechanics, even if we have only a single degree of freedom coupled to a bath, the motion of this system can exhibit collective behavior typically observed in systems with many degrees of freedom. Alternatively, one may also see this as evidence that hydrodynamics is not limited to subsystems with many degrees of freedom; it can also emerge from a single quantum particle \emph{coupled to an environment} at late times, which are larger than the timescale of decoherence. Therefore, the late-time dynamics of the Caldeira-Leggett model should be well described by hydrodynamics.  More evidence to this claim will come from Section \ref{sec: Derivation from Kinetic Theory}, where we discuss how the hydrodynamic behavior of the Caldeira-Leggett model can be derived from an effective kinetic theory for its Wigner function.

\subsection{The evolution of viscous stress and velocity}

In this section, we focus on the dynamical evolution of the velocity $\vec{u}$ and viscous stress $\pi_{ij}$, which can be determined from $\mathcal{C}_{ij}$. First, we  calculate the equilibrium value of $\mathcal{C}_{ij}$, assuming the system reaches an equilibrium steady state where $\partial _t\mathcal{C}_{ij}=0$. At equilibrium, there are no gradients of the flow velocity (considering a non-rotating equilibrium state). Furthermore, the convective terms involving $\vec{u}$ vanish if there is no bulk flow, or they may be treated as small perturbations.
Thus, in equilibrium
\begin{equation}
    \label{Eq:equilibriumstress2}
    \mathcal{C}_{ij}^{\text{eq} }=-n\,k_BT\delta _{ij}\equiv -P\delta_{ij}.
\end{equation}
As discussed previously, $\mathcal{C}_{ij}$ can be interpreted as the stress tensor, which makes $P$ the equilibrium pressure. This identification of $P$ as the equilibrium pressure is naturally supported by the fact that it provides an isotropic equilibrium value for the momentum flux. This reveals a remarkable result: the equilibrium \emph{equation of state} of the Caldeira-Leggett model (a simple quantum particle coupled to a bath of harmonic oscillators at temperature $T$) is the ideal gas equation of state $P=n\,k_B T$.

Now that the equilibrium pressure is determined, we can determine the out-of-equilibrium terms in the stress tensor. This is done by decomposing this tensor as follows
\begin{equation}
    \mathcal{C}_{ij}=\mathcal{C}_{ij}^{\text{eq} }+\pi_{ij}=-P\delta_{ij}+\pi_{ij},
    \label{eq:decomposition of C}
\end{equation}
where $\pi^{ij}$ is the out-of-equilibrium correction to the stress tensor, which we will interpret as a viscous correction later. One may use this decomposition in Eq.~\eqref{eq:momentum equation} to find
\begin{equation}
       \frac{\text{D}}{\text{D}t}  \vec{u} =-\frac{1}{mn} \nabla _{+}P+\frac{1}{mn}\nabla _{+}\cdot \pi  -\frac{1}{m}\nabla _{+}V-2\gamma \vec{u},
\end{equation}
where $\text{D}/\text{D}t =\partial _t+\vec{u}\cdot \nabla _+$ is the comoving (material) derivative. 
The equation of motion for the viscous stress $\pi_{ij}$ is determined using Eq.~\eqref{eq:equation of motion for $C_{ij}$}, which gives
\begin{equation}
    \frac{\text{D}}{\text{D}t} \pi _{ij}+\left ( \nabla _{+}^{k}u_k \right )\pi _{ij}+\left ( \nabla _{+}^{k}u_i \right )\pi _{kj}+ \left ( \nabla _{+}^{k}u_j \right )\pi _{ik}+4\gamma\pi_{ij}=nk_BT\left ( \nabla _{+\,i}u_j+\nabla _{+\,j}u_i  \right ),
\end{equation}
We can rewrite the equation above into a form that is more similar to the transient fluid dynamic theories put forward by Muller \cite{MIS-1}, Israel \cite{MIS-2} and Stewart \cite{MIS-4,MIS-6}
\begin{equation}
    \tau _{\pi }\left [ \frac{\text{D}}{\text{D}t}  \pi _{ij}+\left ( \nabla _{+}^{k}u_k \right )\pi _{ij}+\left ( \nabla _{+}^{k}u_i \right )\pi _{kj}+ \left ( \nabla _{+}^{k}u_j \right )\pi _{ik} \right ] +\pi_{ij}=2\eta \sigma_{ij}+\zeta \delta_{ij} \nabla _{+}^{k}u_k,
    \label{Eq:ISequation}
\end{equation}
where we defined the relaxation time $\tau_\pi$, the shear viscosity $\eta$, the bulk viscosity $\zeta$, and the shear tensor $\sigma$
\begin{equation}
    \tau _\pi=\frac{1}{4\gamma },\quad \eta =\frac{nk_BT}{2} \tau_\pi,\quad \zeta =\frac{4\eta }{3} ,\quad \sigma^{ij}=\left ( \nabla _{+}^iu^j+\nabla _{+}^ju^i -\frac{2}{3}\delta ^{ij}\nabla _{+}^{k}u_k  \right ) .
\end{equation}
Since $\gamma>0$, we note that $\tau_\pi, \eta,\zeta>0$.

This relaxation equation is commonly found in extended irreversible thermodynamics \cite{JouetallBook}, and can be seen as a non-relativistic version of Mueller-Israel-Stewart theories of transient relativistic fluid dynamics \cite{MIS-1,MIS-6}, which are commonly used to investigate the hydrodynamic expansion of the quark-gluon plasma formed in heavy-ion collisions \cite{Romatschke:2017ejr}.  In fact, the presence of terms $\sim (\nabla_+^k u_j) \pi_{ik}$ on the left-hand side of Eq.~\eqref{Eq:ISequation} is a common feature of the DNMR theory of transient relativistic fluid dynamics \cite{Denicol:2012cn}, which is often applied in that context. A distinct difference between transient hydrodynamics and the usual Navier-Stokes equations is that the former have finite characteristic velocities, a hallmark of hyperbolic partial differential equations, while the latter are parabolic (for more information about the mathematical properties of these equations, see \cite{Lerman:2023qyc}). 

Concerning the transport coefficients that appear in Eq.~\eqref{Eq:ISequation}, we note that $\tau_\pi$, $\eta$, and $\zeta$ are entirely determined by the damping constant $\gamma$ and the pressure $P$. This nicely demonstrates the relationship between the dissipative phenomena present in this emerging hydrodynamic theory and the fact that the underlying quantum system is open. Furthermore, it is interesting to notice that 
$\eta/\tau_\pi = P/2$, similar to the results typically found when such transport coefficients are determined from kinetic theory \cite{Denicol:2011fa}. 

Israel-Stewart-like equations, such as Eq.~\eqref{Eq:ISequation}, also display  \textit{ballistic} behavior when the relaxation time $\tau_\pi$ is much larger than the timescales $\tau_S$ under consideration.\footnote{This may be understood as follows. Suppose we have a conserved density $n(x,t)$. Conservation requires that $\partial_t n(x,t) + \nabla \cdot j(x,t) = 0$, where $j(x,t)$ is the associated current. We consider the case where the current itself relaxes on a finite timescale $\tau_\pi$, then we have the Maxwell–Cattaneo relaxation equation $\tau_{\pi} \partial_t j(x,t) + j(x,t) = -D \nabla n(x,t)$. This equation states that the current does not instantaneously become proportional to the gradient of $n$; instead, it relaxes toward Fick’s law over a time $\tau_\pi$. Combining the equations, we have $\tau_{\pi} \partial_t^2 n(x,t) + \partial_t n(x,t) = D \nabla^2 n(x,t)$. Taking 
$\partial_t n \sim 1/\tau_S$, we see that when $\tau_S \gg \tau_\pi$, $\tau_{\pi} \partial_t^2 n \ll \partial_t n$, and the equation reduces to $\partial_t n(x,t) = D \nabla^2 n(x,t)$, which is the usual diffusion equation. In this regime, the current adjusts instantaneously to the gradient, and the behavior is diffusive. However, when $\tau _S\ll \tau _\pi$, the $\tau_{\pi} \partial_t^2 n$ term dominates over the first-order $\partial_t n$ term, and the equation becomes $\tau_{\pi} \partial_t^2 n(x,t) = D \nabla^2 n(x,t)$. This is a wave equation with characteristic velocity $v=\sqrt{D/\tau _\pi }$. In this ballistic regime, perturbations in the density have a finite maximal propagation speed $v$ rather than spreading diffusively.} In this ballistic limit, the particle no longer diffuses, and the viscous stress equation approaches Hooke's law model of elasticity \cite{Gavassino:2023xkt}, which is not typically associated with hydrodynamics. However, the ballistic behavior does not appear in the specific model considered here, once the original assumptions behind the Caldeira-Leggett master equations are employed. Specifically, we work in the high-temperature limit $k_B T \gg \hbar \omega$, assume weak coupling $\hbar \gamma \ll \text{Min} \left \{ \hbar \Lambda, 2\pi k_B T \right \}$, and consider systems with characteristic timescales slower than the bath correlation time, $\hbar \omega_0 \ll \text{Min} \left \{ \hbar \Lambda, 2\pi k_B T \right \}$. These conditions ensure the separation of timescales $\tau_\pi \gg \tau_B$ and $\tau_S \gg \tau_B$. Importantly, when $\tau_\pi \gg \tau_S$, the system operates in the quantum optical limit rather than the quantum Brownian motion regime \cite{10.1093/acprof:oso/9780199213900.001.0001}. For the Caldeira-Leggett model, $\tau_\pi \sim \tau_S$ or $\tau_S \gg \tau_\pi$ is typical, and since we are focused on late-time dynamics, assuming $\tau_S > \tau_\pi$ is reasonable. As a result, these underlying factors preclude ballistic behavior, which confines the dynamics to the hydrodynamic regime.

To determine the leading order behavior of the viscous stress close to equilibrium, one may rewrite Eq.~\eqref{Eq:ISequation} as
\begin{equation}
     \pi_{ij}=2\eta \sigma_{ij}+\zeta \delta_{ij} \nabla _{+}^{k}u_k-\tau _{\pi }\left [ \frac{\text{D}}{\text{D}t}  \pi _{ij}+\left ( \nabla _{+}^{k}u_k \right )\pi _{ij}+\left ( \nabla _{+}^{k}u_i \right )\pi _{kj}+ \left ( \nabla _{+}^{k}u_j \right )\pi _{ik} \right ],
    \label{Eq:ISequationnew}
\end{equation}
and solve this equation for $\pi_{ij}$, iteratively, assuming a derivative expansion \cite{Baier:2007ix}.  To first order in derivatives, one then finds the asymptotic solution
\begin{equation}
    \pi_{ij} = 2\eta \sigma_{ij} +\delta_{ij} \zeta \nabla_+^k u_k  + \mathcal{O}(\partial_t \nabla_+,\nabla_+^2).
    \label{eq:defineNSlimit}
\end{equation}
This gives 
\begin{equation}    
\mathcal{C}_{ij}=-P\delta_{ij}+\pi_{ij}=-P\delta_{ij}+2\eta \sigma_{ij} + \zeta \nabla_+^k u_k \delta_{ij},
\end{equation}
which may be used in Eq.~\eqref{eq:momentum equation} to determine the evolution of the momentum density in this asymptotic regime
\begin{equation}
    \partial _t\vec{u} +\left ( \vec{u}\cdot \nabla _+   \right ) \vec{u} =-\frac{1}{mn} \left [ \nabla_+ P - 2\eta \nabla^2_{+} \vec{u} - \left( \frac{2 \eta}{3} + \zeta \right) \nabla_{+} (\nabla_{+} \cdot \vec{u}) \right ] -\frac{1}{m}\nabla _{+}V-2\gamma \vec{u}  .
    \label{eq:approximateNS}
\end{equation}
The set of equations, Eq.~\eqref{eq:conservationlawdensity} and Eq.~\eqref{eq:approximateNS}, define a compressible Navier-Stokes-like fluid with pressure $P=n k_B T$, nonzero shear and bulk viscosities, in the presence of a source term $\sim \gamma \vec{u}$, which describes how momentum is exchanged with the bath. Therefore, one can see that the late time, long wavelength dynamics of a single quantum particle coupled to a bath of oscillators, \emph{a la} Caldeira-Leggett, is described by Navier-Stokes equations. This unambiguously determines how hydrodynamic behavior emerges in this open quantum system, as the Navier-Stokes constitutive relations are realized in the asymptotic dynamics. 
We will investigate this onset of hydrodynamic behavior in Section \ref{sec: Derivation from Kinetic Theory} using an analytical solution for the exact $\pi_{ij}$ in Caldeira-Leggett theory.

\subsection{Emergence of collective modes}
Understanding the emergence of collective modes is a fundamental aspect of hydrodynamic theory, as it allows us to probe the system's response to small perturbations around equilibrium. In this section, starting from the hydrodynamic equations governing the evolution of density $n$, velocity field $\vec{u}$, and the viscous stress tensor $\pi_{ij}$, we introduce small perturbations around their equilibrium values and determine the linearized equations of motion for these disturbances. In equilibrium, $\pi_{ij}=0$, and for simplicity, we consider a background state of constant density $n_0$ and zero velocity, so that 
\begin{equation}
   n = n_0 + \delta n, \quad  \vec{u} =0+ \delta \vec{u},
   \quad \pi_{ij}=0+\delta \pi_{ij}
\end{equation}
Here, $\delta n$, $\delta \vec{u}$, and $\delta \pi_{ij}$ represent small disturbances around equilibrium\footnote{The temperature of the bath, $T$, is assumed to be constant and uniform.}. The linearized hydrodynamic equations governing the evolution of these perturbations are given by
\begin{subequations}
\begin{align}
    \partial_t \delta n + n_0 \nabla_+^i \delta u_i = 0, \\
    \partial _t \delta u_j = \frac{1}{m n_0} \nabla_+^i \left( -\delta n\, k_B T \delta_{ij} + \delta \pi_{ij} \right) - 2 \gamma \delta u_j, \\
    \partial _t \delta \pi^{ij} +  \frac{1}{\tau_\pi}\delta \pi^{ij} = n_0 k_B T (\nabla_+^i \delta u^j + \nabla_+^j \delta u^i).
\end{align}
\end{subequations}
To solve for the collective modes of the system, we assume plane-wave solutions for the perturbations, where all dynamical quantities are represented as oscillatory functions with a well-defined wavevector $\vec{k}$ and frequency $\omega$. These are given by
\begin{align}
    &\delta n(\vec{x}_+, t) = n' e^{i (\vec{k} \cdot \vec{x}_+ - \omega t)}, &
    & \delta u_j(\vec{x}_+, t) = u'_j e^{i (\vec{k} \cdot \vec{x}_+ - \omega t)}, &
    & \delta \pi_{ij}(\vec{x}_+, t) = \pi'_{ij} e^{i (\vec{k} \cdot \vec{x}_+ - \omega t)}.&
\end{align}
By substituting these plane-wave solutions into the linearized equations, we can determine the dispersion relations for the system and identify the collective modes. We find
\begin{subequations}
\begin{align}
    -i \omega n' + i n_0 k^i u'_i & = 0 , \\
    -i\omega u'_j + \frac{i}{m n_0} k^i \left( k_B T n' \delta_{ij} - \pi'_{ij} \right) + 2 \gamma u'_j & = 0 , \\
    -i\omega \pi'_{ij} + \frac{1}{\tau_\pi} \pi'_{ij} - i \,n_0 k_B T \left( k_i u'_j + k_j u'_i \right) & = 0 .
\end{align}
\end{subequations}
The dispersion relations are obtained by solving this system of equations for $\omega=\omega(k)$. The modes are determined from the following equations
\begin{subequations}
\begin{align}
\label{modetrivial}
    i + \tau_{\pi} \omega  & = 0 \, , \\
    m n_0 \tau_\pi \omega^2 + m n_0 (2\gamma \tau_\pi+1)\omega - n_0 k_B T \tau_\pi k^2  - 2m n_0 \gamma & = 0 \, \label{IV.50b} , \\
    -2 k^2 \omega\, n_0 k_B T \tau_\pi - n_0(i+\omega \tau_\pi)\left[k^2 k_B T - m \omega (2i \gamma + \omega)\right] & = 0
    \label{IV.50c}\, ,
\end{align}    
\end{subequations}
where $v_s^2=k_BT/m$. 
Equation \eqref{modetrivial} is trivially solved, giving $\omega = -i / \tau_\pi = -4i\gamma$, which defines a non-propagating, stable (i.e., $\mathrm{Im}\, \omega$ is non-positive), non-hydrodynamic mode associated with purely transverse (i.e., perpendicular to $\vec{k}$\,) shear disturbances. Equation \eqref{IV.50b} defines longitudinal disturbances given by 
\begin{equation}
    \omega=-3i\gamma \pm \sqrt{v_s^2k^2-\gamma ^2},
    \label{eq:dispersion 2}
\end{equation}
both of which are stable. Taylor expanding these modes around $k=0$, one obtains 
\begin{subequations}
\begin{align}
    \omega & = -2 i \gamma - i D k^2 +\mathcal{O}(k^4), \label{eq:mode1} \\
    \omega & = -4 i \gamma +  i D k^2+\mathcal{O}(k^4). \label{eq:mode2}
\end{align}
\end{subequations}
where above we defined the diffusion coefficient $D=v_s^2/2\gamma$, which depends on the damping constant in the way expected from the Einstein relation, while being also consistent with diffusive behavior known to emerge in quantum Brownian motion. 
For the modes in Eq.~\eqref{eq:mode1} and Eq.~\eqref{eq:mode2}, the first term $-2i\gamma$ represents an \textit{intrinsic} decay rate caused by coupling to the environment, while $-iDk^2$ represents standard diffusive behavior. The exact dispersion Eq.~\eqref{eq:dispersion 2} has square root branch points at $k_{\rm BR}=\gamma/v_s$, which set the radius of convergence for the small-$k$ Taylor series used in Eqs.~\eqref{eq:mode1}-~\eqref{eq:mode2}. For $|k|<k_{\rm BR}$ the modes are strictly damped: $\Im\omega_+=-3\gamma+\sqrt{\gamma^2-v_s^2k^2}\le -2\gamma$ and $\Im\omega_-=-3\gamma-\sqrt{\gamma^2-v_s^2k^2}< -3\gamma$. For $|k|>k_{\rm BR}$ the square root is real, so $\Im\omega_\pm=-3\gamma<0$ and the modes remain damped. The instability suggested by the quadratic truncation of Eq.~\eqref{eq:mode2} arises from solving $-4\gamma + Dk^2=0$ with $D=v_s^2/(2\gamma)$, giving $k_*=\sqrt{8}\gamma/v_s=\sqrt{8}k_{\rm BR}$, which lies outside the series’ convergence disk. Thus the sign change appears only by overextending the low-$\vec{k}$ expansion and there are no truly unstable modes at any $\vec{k}$.

Perturbatively solving Eq.~\eqref{IV.50c} for $\omega$, we obtain in the long-wavelength regime the following transverse modes
\begin{subequations}
\begin{align}
    \omega & = -iDk^2 + \mathcal{O}\left(k^4\right), \label{eq:67a} \\
    \omega & = -2i\gamma - iDk^2 + \mathcal{O}\left(k^4\right), \label{eq:67b} \\
    \omega & = -4i\gamma + 2iDk^2 + \mathcal{O}\left(k^4\right). \label{eq:67c}
\end{align}
\end{subequations}
The first mode in Eq.~\eqref{eq:67a} is the sole hydrodynamic mode of the spectrum. This mode is purely diffusive and stems from the local conservation of probability expressed in Eq.~\eqref{eq:conservationlawdensity}. In contrast, the second and third modes in Eq.~\eqref{eq:67b} and Eq.~\eqref{eq:67c} are non-hydrodynamic modes, which emerge from the coupled momentum and stress equations and reflect the non-conservation of momentum due to the coupling with the environment. We also note that non-hydrodynamic modes, similar to Eq.~\eqref{eq:67b} and Eq.~\eqref{eq:67c}, have been found before in \cite{PhysRevE.97.012130} in the context of spontaneous symmetry breaking in an $O(N)$ scalar field theory obeying a Fokker-Planck equation.  

We remark that the non-hydrodynamic modes found above belong to the class of quasi-hydrodynamic modes discussed in \cite{PhysRevD.99.086012}. This is because on the time scales of interest—namely, those where hydrodynamics serves as a controlled effective theory, these modes decay very slowly. In this regime, we say that momentum is \textit{approximately} conserved. According to the analysis above, we know that when the time scale is much larger than the decoherence time scale $\tau_D$, hydrodynamics can effectively describe the system. On the other hand, the relaxation time of momentum is of order $\tau_P \sim 1/\gamma$ and, therefore, momentum is approximately conserved on this scale $\tau_D \ll t < \tau_P$. This separation of time scales is in fact reasonable because $\tau_D/\tau_P \sim \lambda^{2}_{\text{th}}/\Delta x^2<1$.

\section{Derivation from Kinetic Theory} \label{sec: Derivation from Kinetic Theory}

In 1954, Takabayasi \cite{10.1143/PTP.11.341} employed the method of moments from kinetic theory \cite{grad:1949kinetic} to derive hydrodynamic equations starting with the time-dependent Wigner distribution function for a closed quantum system \cite{PhysRev.40.749}. By integrating the Wigner function and its associated kinetic equations over different powers of momentum, Takabayasi derived continuity and momentum equations that resemble classical hydrodynamics equations. Specifically, the zeroth moment of the Wigner function yields a continuity equation, capturing the conservation of probability density in the quantum fluid. The first moment leads to an equation for momentum flow, which corresponds to the quantum analog of the Euler equation in classical fluid dynamics. This is a standard method in kinetic theory, where the degrees of freedom of a system over short distances are coarse-grained to obtain physical quantities that describe the macroscopic properties of the system. In this section, we show how the moment expansion can be used to characterize the emergence of hydrodynamic behavior in an open quantum system, starting from the evolution equation for the corresponding Wigner function.

The Caldeira-Leggett master equation in coordinate representation can be written as
\begin{equation}
    \frac{\partial \rho}{\partial t} =  \frac{i\hbar}{2m} \frac{\partial^2 \rho}{\partial x_+^i \partial x_-^i} 
- \frac{i}{\hbar} \left[ V(x_+^i + x_-^i) - V(x_+^i - x_-^i) \right] \rho
- 2\gamma x_-^i \frac{\partial \rho}{\partial x_-^i} 
- \frac{8m \gamma k_B T}{\hbar^2} x_-^i x_-^i \rho .
    \label{eq: Density Evolution 2}
\end{equation}
We now define the Wigner function as follows
\begin{equation}
    f( \vec{x} _+,\vec{p},t) = \frac{1}{ \left ( \pi\hbar \right )^3 } \int d \vec{x}_-\,\rho(\vec{x} _+,\vec{x}_-,t)\,e^{-2i\,\vec{p}\cdot \vec{x} _-/\hbar}.
    \label{eq:defineWigner}
\end{equation}
We are interested in the regime where quantum fluctuations are small, which translates into neglecting terms of $\mathcal{O}(\hbar^2)$ in an $\hbar$-expansion, see Appendix \ref{appendix: Caldeira-Leggett Model from Schwinger-Keldysh Formalism}. At this order, the evolution equation for the Wigner function is given by
\begin{equation} 
    \frac{\partial f}{\partial t} +\frac{p^i}{m} \frac{\partial f}{\partial x_+^i} - \frac{\partial V}{\partial x_+^i} \frac{\partial f}{\partial p_i} - 2\gamma \frac{\partial}{\partial p^i}\left ( p^if \right ) -2m\gamma k_B T \delta^{ij} \frac{\partial^2 f}{\partial p^{i}\partial p^j}=0.
    \label{eq:Quantum Klein-Krammers Equation}
\end{equation}
This is the same Fokker-Planck equation found for the Wigner function by Caldeira and Leggett in \cite{Caldeira:1982iu}. 
The terms without $\gamma$ stem from the standard unitary limit, derived from the von Neumann equation. The third term on the right-hand side of Eq.~\eqref{eq:Quantum Klein-Krammers Equation} represents the friction term, which captures the dissipative effects from the environment, leading to momentum damping. The last term in Eq.~\eqref{eq:Quantum Klein-Krammers Equation} leads to momentum diffusion, induced by stochastic forces arising from thermal fluctuations.

\subsection{Moments of the Wigner function}

Next, starting from Eq.~\eqref{eq:Quantum Klein-Krammers Equation}, we use the method of moments to derive the corresponding hydrodynamic equations. We define the zeroth and first moments of the Wigner function as follows
\begin{equation}
    n\left ( \vec{x}_+,t \right ) =\int f\left ( \vec{x}_+,\vec{p},t \right )d^3\vec{p} ,\qquad\qquad mnu^i=\int d^3\vec{p}\, p^if\left ( \vec{x}_+,\vec{p},t \right ).
    \label{eq:zero-th moment of Wigner function}
\end{equation}
Integrating Eq.~\eqref{eq:Quantum Klein-Krammers Equation} over $\vec{p}$, one obtains the continuity equation
\begin{equation}
    \partial _t n+\nabla_+\cdot \left ( n\vec{u}\,  \right )=0 .
\end{equation}
To derive the dynamical equation for the momentum density, we need to define the following second moment of the Wigner function
\begin{equation}
    mn\Pi ^{ij}=\int d^3\vec{p}\, p^ip^j f\left ( \vec{x}_+,\vec{p},t \right ),
\end{equation}
with which one finds
\begin{equation}
    \frac{\partial }{\partial t} \left ( nu^j \right ) +\frac{\partial }{\partial x_{+}^i} \left ( \frac{n}{m}\Pi ^{ij}\right )+ \frac{n}{m}\frac{\partial V}{\partial x_{+}^i}  =-2\gamma nu^j .
\end{equation}
Using the decomposition
\begin{equation}
    \Pi ^{ij} \equiv  mu^iu^j-\frac{1}{n}\mathcal{C}^{ij} ,
    \label{eq:stress tensor here}
\end{equation}
we arrive at 
\begin{equation}
    \frac{\partial u^j}{\partial t} +u^i\frac{\partial u^j}{\partial x_{+}^{i}} =\frac{1}{mn} \frac{\partial \mathcal{C}^{ij}}{\partial x_{+}^{i}}-\frac{1}{m}\frac{\partial V}{\partial x_{+}^{i}} -2\gamma u^j.
\end{equation}
Thus, one recovers Eq.~\eqref{eq:momentum equation} obtained using our position-space decoherence-induced series expansion of the Caldeira-Leggett density matrix in Section \ref{sec: Emergent Viscous Hydrodynamics}.

The evolution equation for $\Pi^{ij}$ can be determined to be the following 
\begin{equation}
    \frac{\partial }{\partial t} \left ( mn\Pi ^{ik} \right )+\frac{1}{m} \frac{\partial}{\partial x_{+}^i}\left ( mnQ^{ijk} \right )  =-\frac{\partial V}{\partial x_{+}^i}\left ( mn u^k \delta_{ij} + mn u^j \delta_{i}^k \right ) -4mn\gamma\Pi^{jk}+4mn\gamma k_BT \delta ^{jk} ,
\end{equation}
where we defined the following third moment of the Wigner function
\begin{equation}
 mnQ^{ijk}=\int d^3\vec{p}\, p^ip^jp^kf\left ( \vec{x}_+,\vec{p},t \right ).
\end{equation}
In order to derive the equation of motion for $\mathcal{C}^{jk}$, we decompose $Q^{ijk}$ in terms of an irreducible rank-3 tensor $B^{ijk}$ as follows
\begin{equation}
Q^{ijk}=m^2u^iu^ju^k+B^{ijk}+\frac{m}{n}\left ( u^i\mathcal{C}^{jk}+u^j\mathcal{C}^{ik}+u^k\mathcal{C}^{ij} \right ).
\end{equation}
One can see that the equation for the rank-2 tensor $\Pi^{ij}$ depends on tensors of lower rank, and also the irreducible rank-3 tensor $B^{ijk}$. Similarly, the equation for the rank-3 tensor will depend on higher-order rank tensors, leading to a hierarchy of coupled partial differential equations for all the moments of the Wigner function (see Appendix \ref{appendix:Moments of Wigner Function}). If no truncation is made, the solution of this infinite series of equations can be formally used to determine the full solution of the Wigner evolution equation in Eq.~\eqref{eq:Quantum Klein-Krammers Equation}. However, in the near-equilibrium, long-time, long wavelength limit, quantities that are not directly related to conserved quantities, such as $B^{ijk}$, become perturbatively small as they encode the contribution from non-hydrodynamic degrees of freedom. Therefore, hydrodynamics corresponds to truncating this infinite series, using some coarse-graining procedure to rewrite higher-order rank tensors in terms of quantities that are naturally related to hydrodynamics, such as the flow velocity, density, etc \cite{Rocha:2023ilf}. The simplest truncation that leads to a transient theory of hydrodynamics consists in setting $B^{ijk}=0$, which leads to the following dynamical equation for $\mathcal{C}^{ij}$
\begin{equation}
    \frac{\partial \mathcal{C}^{jk}}{\partial t}+u^i\frac{\partial \mathcal{C}^{jk}}{\partial x_{+}^i} =-\mathcal{C}^{jk}\frac{\partial u^i}{\partial x_{+}^i}-\mathcal{C}^{ij}\frac{\partial u^k}{\partial x_{+}^i}
    -\mathcal{C}^{ik}\frac{\partial u^j}{\partial x_{+}^i}  -4\gamma \mathcal{C}^{jk}-4\gamma nk_BT\delta ^{jk}.
\end{equation}
We can see that, in equilibrium, $\mathcal{C}^{ij} \to -n k_B T \delta^{ij}$. Therefore, we recover our previous result that the pressure of our effective kinetic system is that of an ideal gas. Writing $\pi^{ij}=\mathcal{C}^{ij}  + n k_B T \delta^{ij} $, we recover Eq.~\eqref{Eq:ISequation} for the evolution of the viscous stress $\pi^{ij}$.

\subsection{Connection to cumulant expansion in kinetic theory}

Now we show that the position-space decoherence-induced series expansion used in Section \ref{sec: Emergent Viscous Hydrodynamics} can be understood in terms of the cumulant expansion for the Wigner function. 
First, because of Eq.~\eqref{eq:defineWigner}, one can see that moments of the Wigner function can be expressed in terms of derivatives of $\rho(\vec{x}_+, \vec{x}_-,t)$. In fact, the moments of $f(\vec{x}_+, \vec{p}, t)$ are given by
\begin{equation}
    \langle p^{i_1} p^{i_2} \cdots p^{i_n} \rangle \equiv \int_{-\infty}^{\infty} d^3\vec{p} \, p^{i_1} p^{i_2} \cdots p^{i_n} f(\vec{x}_+, \vec{p}, t),
\end{equation}
and, using Eq.~\eqref{eq:defineWigner}, one can see that they can also be obtained by  differentiating $\rho(\vec{x}_+, \vec{x}_-)$ with respect to $\vec{x}_-$
\begin{equation}
    \langle p^{i_1} p^{i_2} \cdots p^{i_n} \rangle = \left( \frac{\hbar}{2i} \right)^n 
    \frac{\partial^n \rho(\vec{x}_+, \vec{x}_-, t)}{\partial x_-^{i_1} \partial x_-^{i_2} \cdots \partial x_-^{i_n}} \bigg|_{\vec{x}_- = 0}.
    \label{eq:moment differentiation formula}
\end{equation}
Therefore, by using Eq.~\eqref{eq:moment differentiation formula}, the first and second moments of the Wigner function are given by
\begin{equation}
    \left \langle p^i \right \rangle  =\left ( \frac{\hbar}{2i}  \right ) \frac{\partial \rho }{\partial x_-^i}\bigg|_{\vec{x}_- = 0}=mnu^i,
\end{equation}
and
\begin{equation}
    \left \langle p^ip^j \right \rangle  =\left ( \frac{\hbar}{2i}  \right )^2 \frac{\partial^2 \rho }{\partial x_-^i\partial x_-^j}\bigg|_{\vec{x}_- = 0}=m^2nu^iu^j-m\mathcal{C}^{ij}=mn\Pi^{ij},
\end{equation}
which are precisely what we encountered before in Eq.~\eqref{eq:zero-th moment of Wigner function} and Eq.~\eqref{eq:stress tensor here}. 

The cumulant generating function is the logarithm of the moment generating function, i.e., $\ln \rho(\vec{x}_+, \vec{x}_-,t)$. Expanding the cumulant generating function in powers of $\vec{x}_{-}$, one then finds
\begin{equation}
    \ln \rho(\vec{x}_+, \vec{x}_-,t) = \ln \rho(\vec{x}_+, 0,t) + \frac{2i}{\hbar} \frac{\left \langle \vec{p}  \right \rangle (t)}{n} \cdot \vec{x} _-+\cdots 
    \label{eq:cumulant expansion}
\end{equation}
Therefore, one can see that truncating $\ln \rho(x_+, x_-)$ up to a given order in the $\vec{x}_-$-expansion corresponds to limiting the cumulant expansion of the Wigner function $f$ to a specific order, with higher-order cumulants describing increasingly fine-grained distributions. This connection highlights that our series expansion for $\rho(\vec{x},\vec{x}\,',t)$ presented in \ref{Sec:IV.A} corresponds to a systematic expansion that captures the dominant contributions in the hydrodynamic regime at low momentum $\vec{p}$. A truncation of this series then defines a low-energy effective description, where higher-order powers of $\vec{x}_-$ correspond to high-energy contributions defined at shorter wavelengths, which are outside the hydrodynamic regime and, thus, irrelevant for the infrared dynamics. 

\begin{figure}[H]
    \centering
    \includegraphics[width=0.4\linewidth]{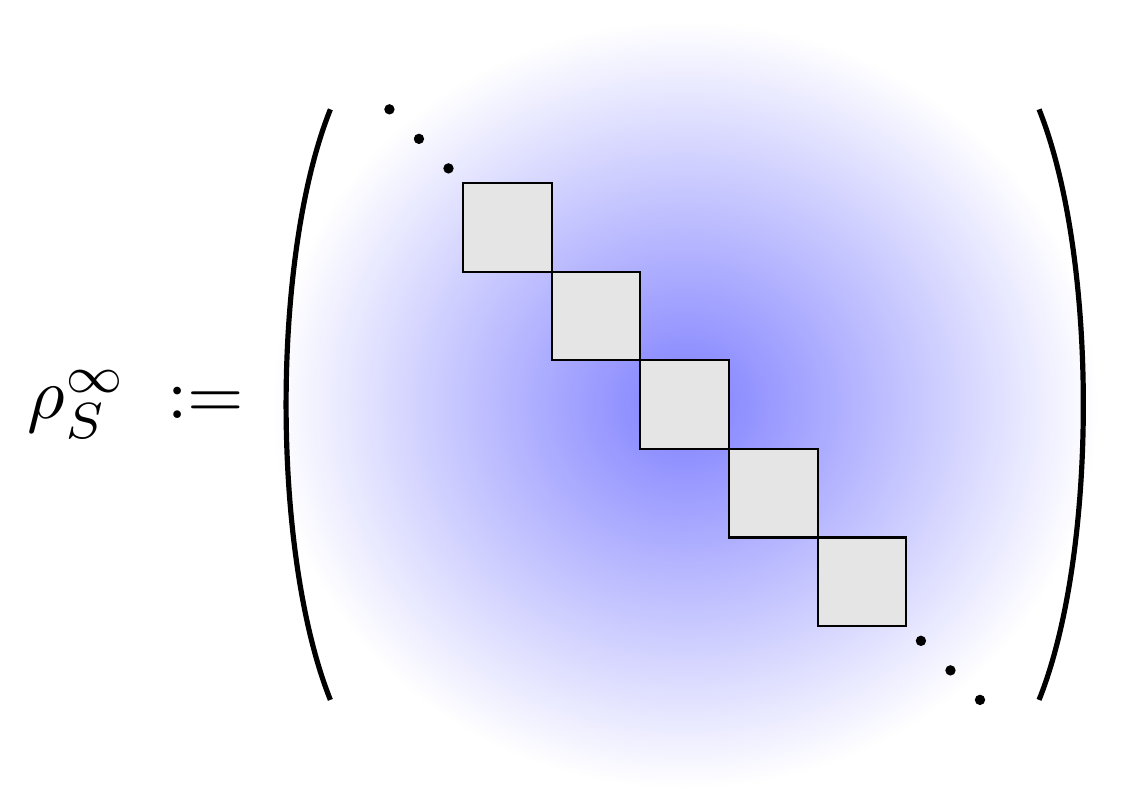}
    \caption{Hydrodynamics of the density matrix. The gray squares represent the diagonal elements, and the purple regions represent the off-diagonal elements. When the density matrix is in the near-equilibrium and semi-classical regime, the dynamics of the diagonal elements and the off-diagonal elements infinitesimally close to the diagonal are well described by hydrodynamics. Decoherence makes the system approximately local in position space, rendering the dynamics far from the diagonal irrelevant. The ensuing hydrodynamic equations provide a complete description of the infrared dynamics of the density matrix.}
    \label{fig:DMhydro}
\end{figure}

The equivalence between the $\vec{x}_{-}$ expansion for the reduced density matrix $\rho_S$ and the cumulant expansion for the Wigner function $f$ may be understood in the following way. By taking the long-time limit of the solution of the Caldeira-Leggett master equation, position-space induced decoherence makes the reduced density matrix of the quantum particle approximately \textit{local} in coordinate space, with deviations from the diagonal elements (i.e., $\vec{x} \neq \vec{x}\,'$) being described by a derivative expansion. This emergent locality ensures the irrelevance of higher-order moments of the Wigner function. These facts, together with the conservation law of probability present in this system, ensure that only low-energy degrees of freedom become relevant in this limit. This is why the reduced density matrix of the system displays hydrodynamic behavior, even though the system in question is just a single quantum particle coupled to an environment -- see Fig.~\ref{fig:DMhydro}.

\subsection{Analytical solution} \label{sec:Analytical Solution}

In this section, we derive an analytical solution to Eq.~\eqref{eq:Quantum Klein-Krammers Equation}, which can be used to investigate the onset of hydrodynamic behavior in the Caldeira-Leggett model. For simplicity, we consider the case where the potential in the Schr\"odinger equation vanishes, i.e., $V=0$. Since Eq.~\eqref{eq:Quantum Klein-Krammers Equation} is a linear equation for $f$, a general solution can be found in terms of the Green's function $G(\vec{x},\vec{x}\,',\vec{p},\vec{p}\,';t)$ as follows
\begin{equation}
f(\vec{x}, \vec{p}, t) = \int d^3\vec{x}\,' \int d^3\vec{p}\,'\,G(\vec{x},\vec{x}\,',\vec{p},\vec{p}\,';t)f_0(\vec{x}\,',\vec{p}\,')
\label{eq:define Greens function}
\end{equation}
where $f_0(\vec{x},\vec{p}\,) =f (\vec{x},\vec{p},0)$ is the initial condition for the Wigner function. The Green's function obeys Eq.~\eqref{eq:Quantum Klein-Krammers Equation} with a delta function $\delta(t)\delta(\vec{x} - \vec{x}\,') \delta(\vec{p} - \vec{p}\,')$ source, so that 
$\lim_{t\to 0^+}G(\vec{x},\vec{x}\,',\vec{p},\vec{p}\,';t) = \delta(\vec{x} - \vec{x}\,') \delta(\vec{p} - \vec{p}\,')$. Using Chandrasekhar's famous analytical result for the solution of the Klein-Kramers equation \cite{RevModPhys.15.1}, one can directly determine that the Green's function is given by
\begin{equation}
    G(\vec{x},\vec{x}\,',\vec{p},\vec{p}\,';t) = \frac{1}{(2\pi \Sigma _X \Sigma_P \sqrt{1-\alpha^2})^3} \exp \left[ -\frac{1}{2(1-\alpha^2)} \left( \frac{|\vec{x} - \vec{\Lambda }_X|^2}{\Sigma_X^2} + \frac{|\vec{p} - \vec{\Lambda}_P|^2}{\Sigma_P^2} - \frac{2\alpha(\vec{x} - \vec{\Lambda}_X) \cdot (\vec{p} - \vec{\Lambda}_P )}{\Sigma_X \Sigma_P} \right) \right],
\end{equation}
where  
\begin{subequations}
    \begin{align}
        & \Sigma_X^2 = \frac{k_B T}{4m \gamma^2} \left[ 1 + 4\gamma t - \left( 2 - e^{-2\gamma t} \right)^2 \right], &
        &\Sigma_P^2 = m k_B T \left( 1 - e^{-4\gamma t} \right),&
        & \alpha  = \frac{k_B T}{2\gamma \Sigma_X \Sigma_P} \left( 1 - e^{-2\gamma t} \right)^2, &
    \end{align}
and 
    \begin{align}
        & \vec{\Lambda } _X = \vec{x}\,' + (2m\gamma)^{-1}(1 - e^{-2\gamma t}) \vec{p}\,', &
        &  \vec{\Lambda } _P = \vec{p}\,' e^{-2\gamma t}. &
    \end{align}
\end{subequations}
Therefore, once the initial condition $f_0(\vec{x},\vec{p}\,)$ is specified, one can use Eq.~\eqref{eq:define Greens function} to determine the solution of the Wigner function and, thus, the expressions for the hydrodynamic variables and any other moment of $f$. 

Here we consider the minimum-uncertainty Gaussian wavepacket as the initial state
\begin{equation}
f_0(\vec{x},\vec{p}\,)=\frac{1}{(\pi\hbar)^3}\;\exp\left({-\frac{(\vec{x}-\vec{x}_0)^2}{\sigma_x^{2}}}\right)\;\exp\left({-\frac{(\vec{p}-\vec{p}_0)^2}{\sigma_p^{2}}}\right),\qquad\sigma_x\sigma_p=\frac{\hbar}{2} .
\end{equation}
We now compute Eq.~\eqref{eq:define Greens function} analytically. For simplicity, we assume $\vec{x}_0, \vec{p}_0=0$, in which case we find the full analytical solution for the Wigner function
\begin{equation} \label{Eq:f_exact}
f(\vec{x}, \vec{p}, t) =
\frac{1}{ ( 2\pi\, \Sigma_{xx}(t) \Sigma_{pp}(t) \sqrt{1 - \rho(t)^2} ) ^3} 
\exp\left[
  -\frac{1}{2(1 - \rho(t)^2)}\left(
    \frac{\vec{x}^2}{\Sigma_{xx}^2(t)}
    + \frac{\vec{p}^2}{\Sigma_{pp}^2(t)}
    - \frac{2\rho(t)\, \vec{x} \cdot \vec{p}}{\Sigma_{xx}(t)\Sigma_{pp}(t)}
  \right)
\right],
\end{equation}
where the covariance matrix is determined by 
\begin{align}
\Sigma_{pp}^2 &= \sigma_p^2 e^{-4\gamma t} +\Sigma_P^2, \quad
\Sigma_{xp}^2 = \frac{\sigma_p^2}{2m\gamma} \left(1 - e^{-2\gamma t} \right) e^{-2\gamma t}+ \alpha \Sigma_X\Sigma_P , \\
\Sigma_{xx}^2 &= \sigma_x^2 + \sigma_p^2 \left({\frac{1 - e^{-2\gamma t}}{2m\gamma } } \right)^2 + \Sigma_X^2.
\end{align}
and $\rho(t) \equiv \Sigma_{xp}^2(t)/\Sigma_{xx}(t)\Sigma_{pp}(t)$. Note that this solution collapses to the Green’s function when the initial packet is taken infinitely sharp 
\begin{equation}
    \lim_{\sigma_x,\sigma_p\to0} f(\vec{x} ,\vec{p},t)=G(\vec{x},\vec{x}\,'=0,\vec{p},\vec{p}\,'=0;t),
\end{equation}
and
\begin{equation}
\lim_{\sigma_x, \sigma_p \to 0} \Sigma_{xx}^2 = \Sigma_X^2, \quad
\lim_{\sigma_x, \sigma_p \to 0} \Sigma_{pp}^2 = \Sigma_P^2, \quad
\lim_{\sigma_x, \sigma_p \to 0} \Sigma_{xp}^2 = \alpha\, \Sigma_X \Sigma_P.
\end{equation}

Using Eq.~\eqref{Eq:f_exact}, we can calculate the \textit{exact} expressions for the hydrodynamic variables and any other moments of the Wigner function. For example, the density is given by  
\begin{equation}
n_{\text{exact}} =  \frac{1}{(2\pi)^{3/2} \Sigma_{xx}^3(t)} \exp\left[ -\frac{|\vec{x}|^2}{2\Sigma_{xx}^2(t)} \right] ,
\end{equation}
while the flow velocity is 
\begin{equation}
u^i_{\text{exact}} = \frac{\rho(t)\, \Sigma_{pp}(t)}{m\, \Sigma_{xx}(t)}\, x^i 
= \frac{\Sigma_{xp}^2(t)}{m\, \Sigma_{xx}^2(t)}\, x^i.
\label{Eq:u_exact}
\end{equation}
One can also determine the second moment of the Wigner function in exact form, which reads
\begin{equation}
\Pi^{ij}_{\text{exact} }= \frac{1}{m} \left[ \Sigma_{pp}^2(t) - \frac{\Sigma_{xp}^4(t)}{\Sigma_{xx}^2(t)} \right] \delta^{ij}
+ \frac{\Sigma_{xp}^4(t)}{m\, \Sigma_{xx}^4(t)}\, x^i x^j.
    \label{eq:exact shear}
\end{equation}
Using the expression above, it is easy to show that
\begin{equation}
\lim_{t \to \infty} \Pi^{ij}_{\text{exact}} \to k_B T \delta^{ij},
\end{equation}
which is precisely what we derived before for the general case. This corresponds to the static equilibrium state, in which both the fluid velocity $\vec{u}$ and the viscous stress $\pi_{ij}$ vanish. The exact expression for the viscous stress is given by
\begin{equation}
        \pi^{ij}_{\text{exact}} = \frac{1}{(2\pi)^{3/2} \Sigma_{xx}^3(t)} \left[ k_B T - \frac{\Sigma_{pp}^2(t)}{m} + \frac{\Sigma_{xp}^4(t)}{m\, \Sigma_{xx}^2(t)} \right]\exp\left[ -\frac{|\vec{x}|^2}{2\Sigma_{xx}^2(t)} \right]  \delta^{ij}.
    \label{eq:exactresultpiji}
\end{equation}
We note that, unlike $\Pi^{ij}$, the viscous stress tensor $\pi^{ij}$ is always isotropic.
It is not difficult to analyze the asymptotic behavior of this analytical expression for $\pi^{ij}$. First, when $t \to \infty$, $\pi^{ij}_{\text{exact}} \to 0$, which means that the viscous stress vanishes as thermal equilibrium is reached, as expected. Furthermore, it can be observed that, as $\gamma$ increases, corresponding to a decrease in the relaxation time $\tau_\pi$, the time required for the viscous stress to asymptote its Navier-Stokes limit becomes progressively shorter. 

It is worth mentioning that we can also verify our previous assertion, namely, $\tau_S > \tau_\pi$, by studying the evolution of $\pi^{ij}_{\text{exact}}$ over time. When $t/\tau_\pi \to 0$ (or, equivalently, $\gamma t \to 0$), 
$\pi^{ij}_{\text{exact}}$ grows quadratically from zero. This reflects our system's initial ballistic motion, which is consistent with the quantum optical limit. Finally, when $t \gg \tau_\pi$, $\pi^{ij}_{\text{exact}}$ decays as a power law, following  $t^{-5/2}$ scaling.

\begin{figure}[t]
    \centering
    \includegraphics[width=0.48\linewidth]{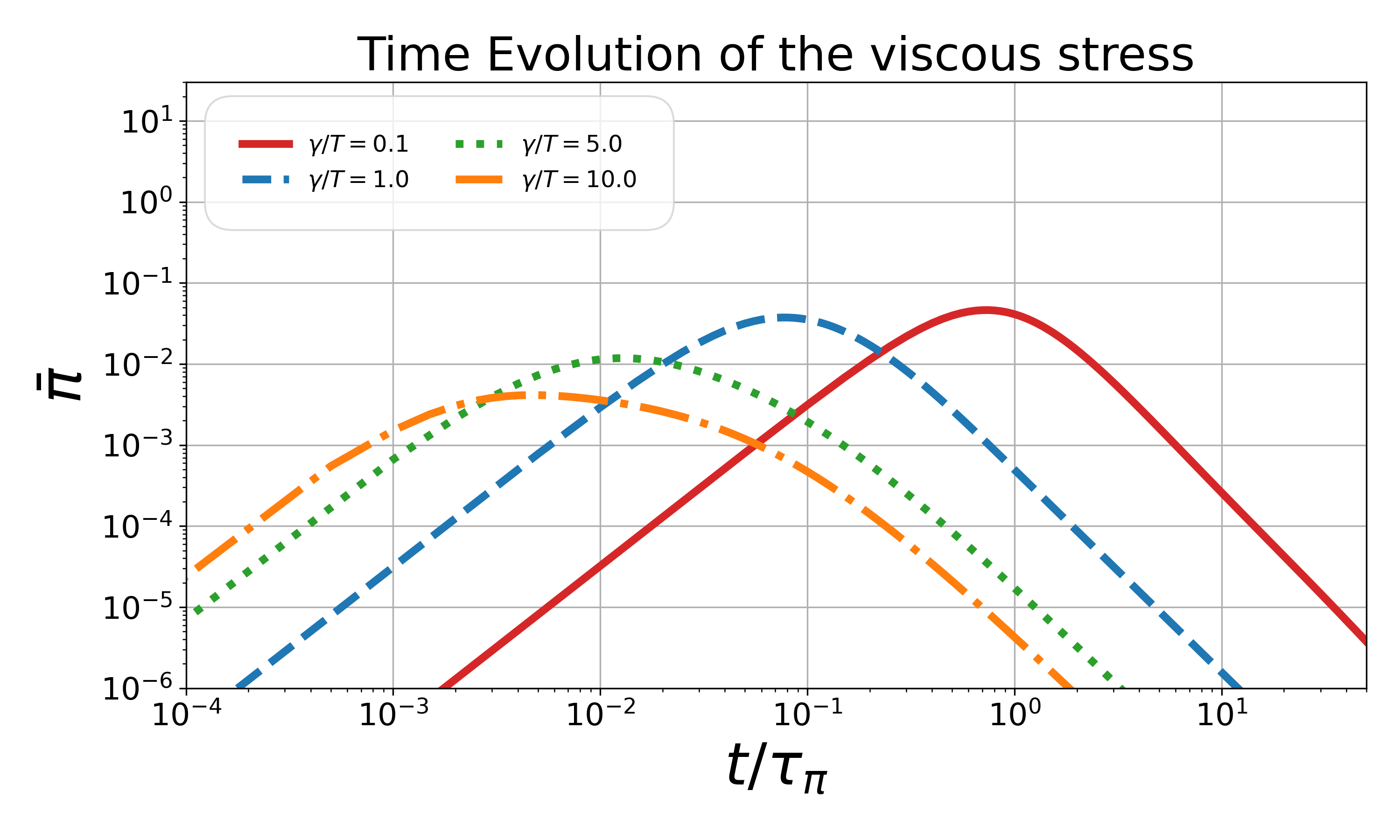}
    \includegraphics[width=0.48\linewidth]{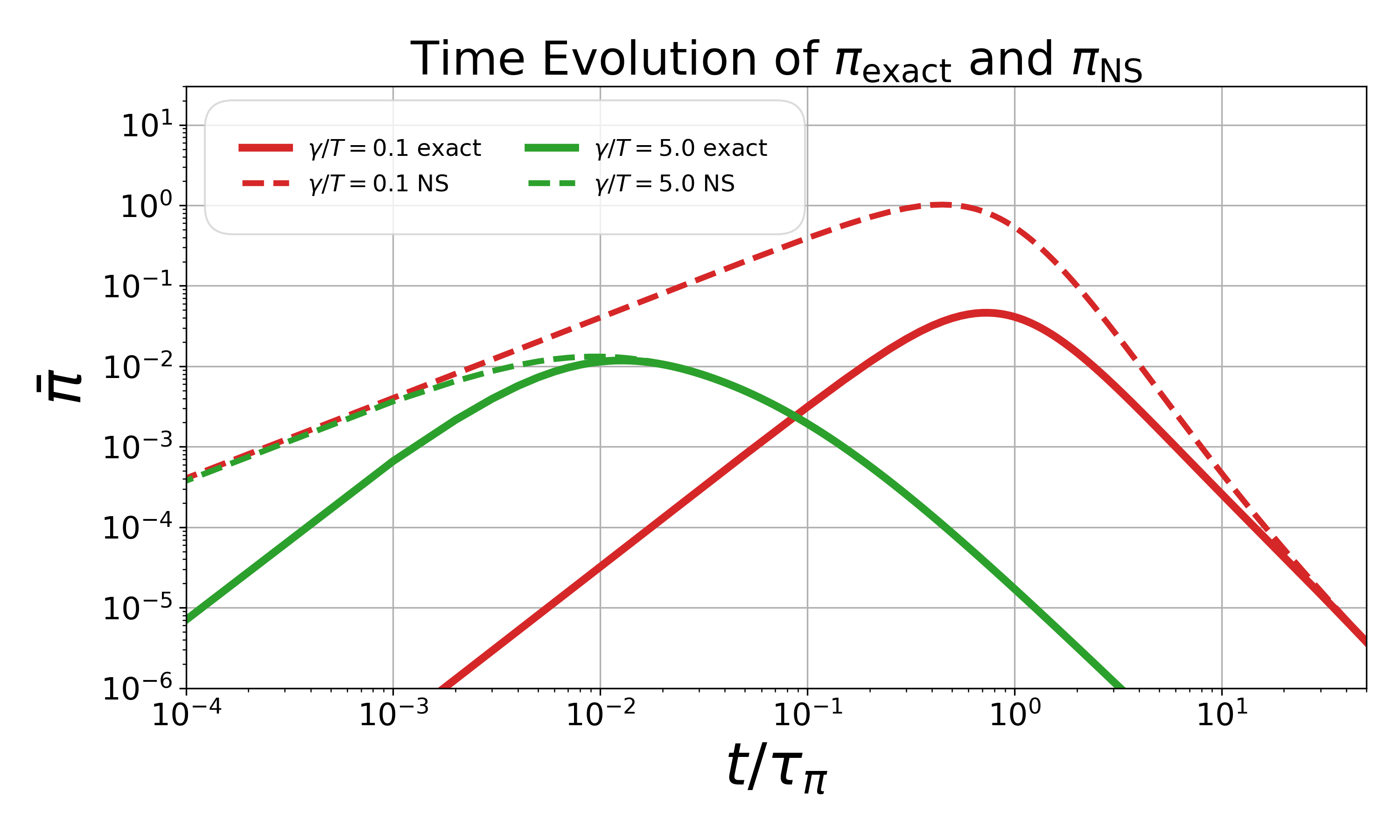}
    \caption{The left panel shows the time evolution for the trace of the viscous stress for different values of $\gamma$ computed using the analytical result in Eq.~\eqref{eq:exactresultpiji}. Here, $\bar{\pi} = \pi / \sqrt{m^3 k_B T \gamma^4}$ is the dimensionless trace of the viscous stress. The curves show how the dimensionless viscous stress approaches equilibrium over time.  The right panel compares the exact solution in Eq.~\eqref{eq:exactresultpiji} and the corresponding Navier-Stokes viscous stress tensor in Eq.~\eqref{eq:exactNS}, for two values of $\gamma$. As expected, the exact solution approaches the Navier-Stokes regime at late times, and this convergence is faster for larger $\gamma$ since $\tau_{\pi} \sim 1/\gamma$.}
    \label{fig:Shear}
\end{figure}

The evolution of the trace $\pi_{\text{exact}}=\delta_{ij}\pi^{ij}_{\text{exact}}/3$ as a function of time, determined using the analytical result in Eq.~\eqref{eq:exactresultpiji}, is shown in Fig.~\ref{fig:Shear}.
The curves on the left-hand side show how the dimensionless viscous stress approaches equilibrium over time, for different values of $\gamma$. One can see that the viscous stress approaches zero at late times, when the system returns to equilibrium.

To find the onset of hydrodynamic behavior in this system, one can use our exact analytical solution and compare it to its corresponding Navier-Stokes expression $\pi_{\text{NS}}^{ij} = 2 \eta \sigma^{ij} + \zeta \nabla_k u^k \delta^{ij}$.
Using the fluid velocity for the exact solution defined in Eq.~\eqref{Eq:u_exact}, we find that the corresponding $\pi_{\text{NS}}^{ij}$ is given by
\begin{equation}
\pi^{ij}_{\text{NS}}(\vec{x}, t) = 
\frac{1 - e^{-2\gamma t}}{(2\pi)^{3/2} \Sigma_{xx}^5(t)}  \, \frac{(k_B T)^2}{4 \gamma^2 m} 
 \exp\left[ -\frac{|\vec{x}|^2}{2\Sigma_{xx}^2(t)} \right]\delta^{ij}.
\label{eq:exactNS}
\end{equation}
Here, it is useful to make a choice for the variances of the initial state. If we select $\sigma_p^2 = m k_B T$ and $\sigma_x^2 = \hbar^2 / (4 m k_B T)$, the exact viscous stress can be related to the Navier-Stokes viscous stress by the simple relation
\begin{equation}
    \pi_{\mathrm{exact}}^{ij}(\vec{x},t) = \left( 1-e^{-2\gamma t} \right) \pi_{\mathrm{NS}}^{ij}(\vec{x},t) .
\end{equation}
This allows us to immediately see that the exact solution approaches the Navier-Stokes result with a rate that scales with the relaxation time $\tau_{\pi}$. Indeed, the plot on the right-hand side of Fig.\ \ref{fig:Shear} shows a comparison of the exact solution in Eq.~\eqref{eq:exactresultpiji} and the corresponding Navier-Stokes viscous stress tensor in Eq.~\eqref{eq:exactNS}, for two values of $\gamma$. When these expressions match, one may say that the Navier-Stokes constitutive relations become a good approximation to the underlying system's out-of-equilibrium dynamics. 
One can see that the exact solution approaches the Navier-Stokes regime at late times, as expected, and this convergence is faster for larger $\gamma$ since $\tau_{\pi} \sim 1/\gamma$. Therefore, as we discussed earlier, only when the considered timescale of the system is greater than the relaxation time does the system exhibit hydrodynamic behavior.

\section{Conclusions}
\label{Conclusion}

In this paper, we have demonstrated explicitly how viscous hydrodynamics can emerge from the quantum Brownian motion of a single particle interacting with a thermal environment, as described by the Caldeira-Leggett model. In principle, all information about this open quantum system is contained in its reduced density operator. Due to decoherence, far-off-diagonal elements of the reduced density matrix in coordinate space decay rapidly. Therefore, to determine the long-time, long wavelength properties of this system, one can focus solely on the dynamical behavior of elements near the diagonal, for which $\vec{x}_{-} = \vec{x} - \vec{x}\,'$ is small ---that is, comparable to the thermal de Broglie wavelength $\lambda_{\textrm{th}}$. As equilibrium is reached, these elements approach a Gaussian of width $\sqrt{2} \lambda_{\textrm{th}}$ around the diagonal \cite{10.1093/acprof:oso/9780199213900.001.0001}. 
Motivated by this position-space decoherence, we introduce a series expansion scheme in powers of $\vec x_-$, which when truncated to second order recovers a Gaussian distribution around the diagonal. 
Performing such a series expansion amounts to disentangling this information into IR and UV components. From the perspective of the Schwinger–Keldysh formalism, the physical meaning of this expansion becomes particularly transparent: one concentrates on the coarse-grained dynamics governed by $\vec{x}_+$ while neglecting the higher-order quantum fluctuations encoded in $\vec{x}_-$. 
We thus  obtain equations describing the dynamics of the diagonal and near-diagonal elements of the reduced density matrix, which 
when combined with the local conservation of probability, naturally leads to the emergence of hydrodynamics, even from a system with only a single degree of freedom. 

Following the expansion outlined above, we obtain an infinite set of dynamical equations describing the evolution of the Caldeira-Leggett system,  reminiscent of the BBGKY hierarchy.
Upon truncating this expansion at second order, we find a set of transient hydrodynamic equations analogous to the non-relativistic limit of the fluid dynamical equations used to describe the quark-gluon plasma formed in heavy-ion collisions. Such equations describe the evolution of the probability density $n$, the fluid velocity $\vec{u}$, and the stress tensor $\mathcal{C}^{ij}$ as dynamical degrees of freedom obeying a set of coupled differential equations. 
The transport coefficients are determined by the temperature of the bath $T$, the probability density, and the damping constant $\gamma$, which characterizes the strength of interactions with the environment. At asymptotically late times, the stress tensor becomes isotropic with a pressure given by the ideal gas equation of state. The relaxation time $\tau_{\pi}$ for viscous corrections to the stress tensor scales as $\tau_{\pi} \sim 1/\gamma$, while the shear $\eta$ and bulk $\zeta$ viscosities scale like $\eta, \zeta \sim P \tau_{\pi}$.

The emergence of viscosity in our system is driven by decoherence due to frequent interactions with a large environment, rather than the large number of interactions among particles in a conventional fluid, which decreases the mean free path, leading to a large separation of length scales traditionally understood as the origin of hydrodynamics. Although our equations of motion exhibit ballistic behavior at early times, such scales are outside the regime of applicability of the Caldeira-Leggett master equation. When interactions with the bath are sufficiently frequent, that is $\gamma \gg \omega$ for frequency $\omega$, the transient equations obtained in this work reduce to the famous Navier-Stokes equations, with an additional Ohm-like drag force encoding the exchange of momentum with the bath. 

In particular, the large number of environmental degrees of freedom contained in the oscillator bath is essential for generating irreversibility, dissipation, and well-defined transport coefficients in our construction. Our analysis therefore should not be interpreted as showing that an isolated quantum system with only a few degrees of freedom can hydrodynamize on its own. Rather, it demonstrates that, once such a large bath is present, the late-time dynamics of single quantum particle observables can be captured by standard viscous hydrodynamic equations, even though the system itself carries only few degrees of freedom. We are currently investigating how these conclusions are modified when the bath itself has only a finite number of degrees of freedom, but this question lies beyond the scope of the present work.

We note that this is very different than the original Madelung result for a pure state, which provides an exact mapping between the Schr\"odinger equation and the ideal Euler equations with a quantum potential in Eq.~(7) valid at all times. In contrast, in our approach, decoherence leads to the emergence of viscous hydrodynamic behavior at late times when the system is near equilibrium.

An alternative way to understand this result follows from an effective kinetic theory formulation of the Caldeira-Leggett model. This can be done by applying the method of moments and focusing on the low-order moments of the Wigner distribution function, defined from the system's reduced density matrix. This procedure leads to the same transient hydrodynamic equations as those discussed above. The correspondence between these approaches implies that our $\vec{x}_{-}$ expansion scheme for the density matrix can be mapped onto a cumulant expansion of the Wigner function, as expressed in Eq.~\eqref{eq:cumulant expansion}. It further illustrates that when the system’s late-time behavior is considered, high-frequency modes have already decayed, and the relevant variables governing the dynamics are the slow degrees of freedom and low-frequency responses, which offers new insights into the universality of hydrodynamics. Finally, using an exact analytical solution of the Wigner function for the Caldeira-Leggett system, we further validate that the late-time dynamics is accurately captured by the Navier-Stokes constitutive relations. This is done by comparing the exact solution for the second moment of the Wigner function to the predicted Navier-Stokes result, shown in Fig.~\ref{fig:Shear}.

Even though our system corresponds to a single quantum particle coupled to a thermal bath, our work may provide some insight into related problems involving the quark-gluon plasma formed in heavy-ion collisions. In fact, transient fluid dynamical equations of Mueller-Israel–Stewart type, similar in spirit to the ones found here, are widely employed in simulations of the quark–gluon plasma. These equations are often derived from relativistic kinetic theory \cite{Rocha:2023ilf}, and our new approach to the density matrix via a cumulant expansion in the Wigner representation also suggests a close relationship to kinetic theory derivations. From a broader physics perspective, our results show that hallmarks of collective behavior -- such as viscosity and hydrodynamic‐like flow -- can emerge in single‐particle observables due to decoherence, which renders the reduced density matrix near diagonal in position space. In particular, one may speculate \cite{Castorina:2007eb} that a nontrivial quantum vacuum could play a role analogous to that of the thermal bath considered in this work. If true, this may help explain why hydrodynamics provides a good description of the spacetime evolution of the matter formed even in systems as small as proton-nucleus or proton-proton collisions, which challenge our traditional understanding of hydrodynamics.

As discussed earlier, our motivation for performing the series expansion comes from position-space decoherence. One may wonder whether this hydrodynamic behavior is a general property of quantum Brownian motion. The first issue one needs to discuss is whether position-space decoherence always occurs within such a system. Specifically, whether the density matrix becomes near-diagonal when the timescale is much larger than the decoherence timescale $\tau_D$. Generally speaking, the answer to this question depends on the form of the system's Hamiltonian and that of the system-environment interaction. Intuitively, terms in either of those, which depend on the system's coordinate $\vec{x}$, will tend to push the reduced density matrix $\rho_S$ towards diagonal form. In contrast, those dependent on the conjugate momentum variable $\vec{p}$ will tend to delocalize it. So, the behavior of $\rho_S$ as a function of time, and in particular its $t\gg \tau_D$ behavior, will result from the competition of these two effects. Therefore, in the case of weak position diffusion, when the timescale is much greater than the decoherence timescale, one can always obtain a near-diagonal density matrix, allowing for a series expansion. Thus, in this case, in the presence of a local conservation law, the late-time dynamics of the system should also be well described by hydrodynamics. A direct extension of our work is to consider the celebrated Hu-Paz-Zhang (HPZ) master equation \cite{PhysRevD.45.2843}, which is an exact master equation describing the Ohmic, sub-Ohmic, or super-Ohmic regimes of quantum Brownian motion of a particle linearly coupled to a thermal bath. Since this equation still exhibits spatial-position decoherence, we expect that its late-time dynamics can likewise be described by hydrodynamics. However, due to the influence of quantum fluctuations at low temperature and non-Markovian effects, the resulting hydrodynamic modes may become subdiffusive or superdiffusive rather than purely diffusive. Also, the resulting truncated hydrodynamic equations may not be of transient type, as found here.

It would be interesting to investigate how the properties of quantum Brownian motion discussed here could be observed experimentally, especially in the context of diffusion of heavy flavor in the quark-gluon plasma \cite{Moore:2004tg}. For example, the Lindblad master equation has already been used to simulate the dynamical behavior of quarkonium in heavy-ion collision experiments \cite{PhysRevD.101.034011, DeJong:2020riy, Yao_2021, hitschfeld2024recentdevelopmentsquarkoniumopen}. Therefore, if these equations satisfy the conditions we have proposed above, the late-time dynamics of quarkonium can be well described by hydrodynamics, with new effective shear and bulk viscosities dictated by the coupling of the quarkonium to the underlying quark-gluon plasma.

Within the paradigm of effective field theory, hydrodynamic modes may be understood as Nambu-Goldstone modes arising from the spontaneous symmetry breaking of continuous symmetries \cite{Dubovsky:2011sj}. In this context, it is interesting to note the new class of spontaneous symmetry-breaking phenomena introduced in \cite{Lee_2023, Ogunnaike_2023}. In fact, spontaneous symmetry breaking was identified as a general mechanism underlying the emergence of hydrodynamics in open quantum systems \cite{Ogunnaike_2023}. This mechanism, known as strong-to-weak spontaneous symmetry breaking, has sparked considerable recent interest, see \cite{Lee_2023, Ogunnaike_2023, Akyuz2024, Delacretaz2025, kim2024errorthresholdsykcodes, lessa2024strongtoweakspontaneoussymmetrybreaking, gu2024spontaneoussymmetrybreakingopen, PhysRevB.110.155150, huang2024hydrodynamicseffectivefieldtheory}. Therefore, it would be interesting to investigate how our results concerning the hydrodynamic behavior of the Caldeira-Leggett model -- a cornerstone of quantum dissipative dynamics -- may be understood from the perspective of strong-to-weak spontaneous symmetry breaking. In that regard, the extension to Lorentz invariant systems may also be pursued.

\section{Acknowledgments}
We thank J.Y. Lee for insightful discussions regarding gapless modes in open quantum systems. J.N. and M.H. thank A.~Kodumagulla for collaborating during the earliest
stages of this work. Z.Z. gratefully acknowledges the support provided by the Tan Family Education Foundation Scholarship and the Illinois Engineering Outstanding Scholarship. N.M. and J.N. are partly supported by the U.S. Department of Energy, Office of Science, Office for Nuclear Physics under Award No. DE-SC0023861. M.H. was partially supported by Universidade Estadual do Rio de Janeiro, within the Programa de Apoio à Docência (PAPD).

\appendix

\section{Caldeira-Leggett Model from the Schwinger-Keldysh Formalism}
\label{appendix: Caldeira-Leggett Model from Schwinger-Keldysh Formalism}

The purpose of this appendix is to derive the effective action of the Caldeira-Leggett model within the framework of the Schwinger-Keldysh (SK) formalism \cite{Schwinger:1960qe, Keldysh:1964ud}. At first glance, this may seem trivial, as the influence functional formalism already yields the effective action. However, starting from the more general Schwinger-Keldysh formalism is useful as it may clarify some conceptual points. 

Let the system be described by one degree of freedom $x(t)$, and the action of the system is given by $S_{\text{sys}}[x] = \int_{t_i}^{t_f} dt \left[ \frac{1}{2}m\dot{x}^2 - V(x) \right]
$. The environment is modeled as $N$ harmonic oscillators with coordinates $x_n(t)$. Each oscillator has mass $m_n$ and frequency $\omega_n$, the bath action is $S_{\text{bath}}[\{x_n\}] = \sum_{n=1}^{N} \int_{t_i}^{t_f} dt \left[ \frac{1}{2} m_n \dot{x}_n^2 - \frac{1}{2} m_n \omega_n^2 x_n^2 \right]$. We also assume the system couples linearly to each bath oscillator, so the interaction action is $S_{\text{int}} = - \int_{t_i}^{t_f} dt \sum_{n=1}^{N} c_n x(t) x_n(t)$. Thus, the total action is
\begin{equation}
    S_{\text{tot}}[x, \{x_n\}] = S_{\text{sys}}[x] + S_{\text{bath}}[\{x_n\}] + S_{\text{int}}[x, \{x_n\}].
\end{equation}
In the SK formalism, we must double the fields for forward and backward time evolution. Denote $x(t)$ and $x'(t)$ for the system coordinate, and $x_n(t)$ and $x_n'(t)$ for each bath oscillator. The SK action is given by
\begin{equation}
    S_{\text{SK}} = S_{\text{sys}}[x] + S_{\text{bath}}[\{x_n\}] + S_{\text{int}}[x, \{x_n\}] 
- \big( S_{\text{sys}}[x\,'] + S_{\text{bath}}[\{x_n'\}] + S_{\text{int}}[x\,', \{x_n'\}] \big) .
\end{equation}
We consider the total path integral over all fields (system + bath), typically with some initial density matrix $\rho(t_i)$. Symbolically
\begin{equation}
    Z = \int \mathcal{D}x \mathcal{D}x\,' \prod_{n=1}^{N} \left( \mathcal{D}x_{n} \mathcal{D}x_{n}' \right)
\rho\left[x(t_i), x\,'(t_i), \{x_{n}(t_i)\}, \{x_{n}'(t_i)\}\right] e^{i S_{\text{SK}}}.
\end{equation}
However, since we are only concerned with the dynamical behavior of the system, we integrate out the bath degrees of freedom $x_n$, $x_n'$, to obtain the effective action for $x$ and $x\,'$ that describes the system's dynamics. Then,
\begin{equation}
    Z = \int \mathcal{D}x \mathcal{D}x\,' \left[ \prod_{n=1}^{N} \int \mathcal{D}x_{n} \mathcal{D}x_{n}' \, \rho[\dots] e^{i S_{\text{SK}}[\dots]} \right] \equiv \int \mathcal{D}x \mathcal{D}x\,'\rho_S[x(t_i),x'(t_i)]\, e^{i\left \{ S_{\text{sys}}[x]-S_{\text{sys}}[x\,']+S_{\text{IF}}\left [ x,x\,' \right ] \right \} },
\end{equation}
where we define the influence functional $\exp\{i S_{\text{IF}}[x, x']\}=\mathcal{F}[x, x\,']$ as the result of integrating out the bath, and $S_{\text{IF}}\left [ x,x\,' \right ]$ is the influence action. The effective SK action for the system alone then becomes
\begin{equation}
    S_{\text{SK}}^{(\text{eff})}[x, x\,'] = S_{\text{sys}}[x] - S_{\text{sys}}[x\,'] + S_{\text{IF}}[x, x\,'].
\end{equation}
Now, the question is to compute the form of $S_{\text{IF}}[x, x\,']$. For the Caldeira-Leggett model, the effective action is given by \cite{Caldeira:1982iu,feynman2010quantum, Calzetta:2008iqa}
\begin{equation}
\begin{split}
        S_{\text{SK}}^{(\text{eff})}[x, x\,'] &= \int dt \left[ \frac{1}{2} m (\dot{x}^2 - \dot{x}\,'^2) - \left( V(x) - V(x\,') \right) \right] + \int dt\, dt\,' \Big\{ \alpha_R(t - t\,') \big[x(t) - x\,'(t)\big] \big[x(t\,') + x\,'(t\,')\big] \Big\}\\
    &+i \int dt\, dt' \Big\{ \alpha_I(t - t\,') \big[x(t) - x\,'(t)\big] \big[x(t\,') - x\,'(t\,')\big] \Big\},
\end{split}
\end{equation}
where $\alpha_R$ and $\alpha_I$ are real and imaginary kernels encoding dissipation and fluctuations, respectively. To clarify the physical meaning of our proposed expansion of the density matrix in terms of $x-x\,'$, we will use the Keldysh basis, which is realized by performing the Keldysh rotation
\begin{equation}
    x_+ = \frac{x +x\,'}{2}, \quad x_- = \frac{x - x\,'}{2}, 
\end{equation}
where $x_{+}(t)$ is referred to as the classical component and $x_{-}(t)$ is referred to as the quantum component. Above, $x_{+}(t)$ represents the average of the forward and backward paths, describing the coarse-grained dynamics, while $x_{-}(t)$ represents the difference between the forward and backward paths, which encodes \textit{quantum fluctuations}. To illustrate why $x_-(t)$ can describe quantum fluctuations, we can write the effective action in its general form in the Keldysh basis as follows
\begin{equation}
    S_{\text{SK}}^{(\text{eff})}[x_{\mathrm{+}}, x_{\mathrm{-}}] = \int dt \left[ \underbrace{\mathcal{A}[x_{\mathrm{+}}] x_{\mathrm{-}}(t)}_{\text{linear in } x_{\mathrm{-}}} + \underbrace{\mathcal{B}[x_{\mathrm{+}}] x_{\mathrm{-}}^2(t)}_{\text{quadratic (fluctuations)}} + \dots \right],
\end{equation}
where $\mathcal{A}$ and $\mathcal{B}$ are functionals determined by expanding $S_{\text{sys}}[x] - S_{\text{sys}}[x\,']$ around $x_+$ and $x_-$. As we discussed earlier, quantum decoherence occurs due to the interaction between the system and the environment, causing the density matrix to gradually approach a diagonal form. This implies that during the dynamical evolution $x_- \to 0$, so we can neglect the higher-order terms of $x_-$ in the effective action, which means we only need to focus on very small fluctuations. Therefore, the physical meaning of $x_- \to 0$ is now quite clear. Namely, when $x_- \to 0$, we are effectively in the semi-classical, near-equilibrium regime. Therefore, it makes sense that the hydrodynamic regime, discussed in this work, emerges when $x_-$ is small.

\section{Moments of Wigner function}
\label{appendix:Moments of Wigner Function}

Here we give the details of the derivation of the evolution equations for the moments of the distribution function using the Wigner equation. 
For instance, the zeroth and first moments of the distribution function are defined by
\begin{equation}
    n\left ( \vec{x}_+,t \right ) =\int f\left ( \vec{x}_+,\vec{p},t \right )d^3\vec{p} ,
\end{equation}
and
\begin{equation}
    mnu^i=\int p^if\left ( \vec{x}_+,\vec{p},t \right ) d^3\vec{p}. 
\end{equation}
We also define the second moment of the Wigner function as
\begin{equation}
    mn\Pi ^{ij}=\int d^3\vec{p}\, p^ip^jf\left ( \vec{x}_+,\vec{p},t \right ).
    \label{eq:Second moments}
\end{equation}

\subsection{Equation of motion for the zeroth moment}

Integrating the Wigner equation, Eq.~\eqref{eq:Quantum Klein-Krammers Equation}, over $\vec{p}$, we obtain
\begin{equation}
    \int \left ( \frac{\partial f}{\partial t} + \frac{p_i}{m} \frac{\partial f}{\partial x_{+}^{i}} - V^{(1)}(x) \frac{\partial f}{\partial p_i} \right )d^3\vec{p}   = 2\gamma \int \left[ \frac{\partial}{\partial p_i} (p_i f) + m k_B T \frac{\partial^2 f}{\partial p_i \partial p_i} \right]d^3\vec{p} .
\end{equation}
Then, using the following properties
\begin{subequations}
\begin{align}
    \int V^{(1) }\frac{\partial f}{\partial p_i}d^3\vec{p}=V^{(1) }\int \frac{\partial f}{\partial p_i}d^3\vec{p}
=V^{(1) }\left [ f\Big|_{-\infty }^{\infty }  \right ] & =0, \\
\int d^3\vec{p}\,\frac{\partial }{\partial p_i} \left ( p_if \right ) =\left ( p_if \right ) \Big|^{\infty }_{-\infty } & =0, \\
\int d^3\vec{p}\, \frac{\partial ^2}{\partial ^2p_i} f=\left ( \frac{\partial f}{\partial p_i}  \right )\Big|^{\infty }_{-\infty } & =0 ,
\end{align}
\end{subequations}
we find 
\begin{equation}
    \frac{\partial }{\partial t} \int fd^3\vec{p} =-\frac{1}{m} \frac{\partial }{\partial x^{i}_{+}} \int p^ifd^3\vec{p} .
\end{equation}
Substituting the definition of the zeroth and first moments, the equation of local probability conservation is obtained
\begin{equation}
    \partial _t n+\nabla_+\cdot \left ( n\vec{u}  \right )=0 .
\end{equation}

\subsection{Equation of motion for the  first moment}

Multiplying the Wigner equation by $p_j$ and integrating over $\vec{p}$, we find that
\begin{equation}
    \int p_j\left ( \frac{\partial f}{\partial t} + \frac{p_i}{m} \frac{\partial f}{\partial x_{+}^{i}} - V^{(1)}(x) \frac{\partial f}{\partial p_i} \right )d^3\vec{p}   = 2\gamma \int p_j\left[ \frac{\partial}{\partial p_i} (p_i f) + m k_B T \frac{\partial^2 f}{\partial p_i \partial p_i} \right]d^3\vec{p} .
\end{equation}
Using 
\begin{subequations}
\begin{align}
    f \to 0 \text{ as } |\vec{p}| & \to \infty, \\
    \int d^3 \vec{p}\,p_j \frac{\partial^2 f}{\partial p_i \partial p_i} & = 0, \\
    \int d^3 \vec{p} \, p_j \frac{\partial}{\partial p_i} \left( p_i f \right) & = - \int d^3 \vec{p} \, p_j f ,
\end{align}
\end{subequations}
we find
\begin{equation}
    \label{eq:B10}
    \frac{\partial }{\partial t} \left ( nu^j \right ) +\frac{\partial }{\partial x_{+}^i} \left ( \frac{n}{m}\Pi ^{ij}\right )+ \frac{n}{m}\frac{\partial V}{\partial x_{+}^i}  =-2\gamma nu^j .
\end{equation}
Now, let us define
\begin{equation}
    \Pi ^{ij} \equiv  mu^iu^j-\frac{1}{n}\mathcal{C}^{ij} .
    \label{eq:stress tensor}
\end{equation}
Substituting this equation into \eqref{eq:B10}, we arrive at Eq.~\eqref{eq:momentum equation}
\begin{equation}
    \frac{\partial u^j}{\partial t} +u^i\frac{\partial u^j}{\partial x_{+}^{i}} =\frac{1}{mn} \frac{\partial \mathcal{C}^{ij}}{\partial x_{+}^{i}}-\frac{1}{m}\frac{\partial V}{\partial x_{+}^{i}} -2\gamma u^j .
\end{equation}

\subsection{Equation of motion for the  second moment}

Again, the equations for the moments have a hierarchical structure with the equations for the $n^{\mathrm{th}}$ moment depending on the $(n+1)^{\mathrm{th}}$ moment. So, to obtain an equation for the second moment, we define the third moment of the distribution function
\begin{equation}
    mnQ^{ijk}=\int d^3\vec{p}\, p^ip^jp^kf\left ( \vec{x}_+,\vec{p},t \right ).
\end{equation}
Multiplying the Wigner equation by $p_jp_k$ and integrating over $\vec{p}$ one finds
\begin{equation}
    \int p_jp_k\left ( \frac{\partial f}{\partial t} + \frac{p_i}{m} \frac{\partial f}{\partial x_{+}^{i}} - V^{(1)}(x) \frac{\partial f}{\partial p_i} \right )d^3\vec{p}   = 2\gamma \int p_jp_k\left[ \frac{\partial}{\partial p_i} (p_i f) + m k_B T \frac{\partial^2 f}{\partial p_i \partial p_i} \right]d^3\vec{p} .
\end{equation}
This equation is equivalent to
\begin{equation}
    \frac{\partial }{\partial t} \left ( mn\Pi ^{ik} \right )+\frac{1}{m} \frac{\partial}{\partial x_{+}^i}\left ( mnQ^{ijk} \right )  =-\frac{\partial V}{\partial x_{+}^i}\left ( mn u^k \delta_{ij} + mn u^j \delta_{ik} \right ) -4mn\gamma\Pi^{jk}+4mn\gamma k_BT \delta ^{jk} .
    \label{eq:evolution of second moment}
\end{equation}
It is convenient to expand out $\partial _t\left ( mn\Pi ^{ik} \right )$ to obtain
\begin{equation}
    \frac{\partial}{\partial t} \left( mn \Pi^{jk} \right) = m \left( \frac{\partial n}{\partial t} \Pi^{jk} + n \frac{\partial \Pi^{jk}}{\partial t} \right)
    = m \left( - \frac{\partial}{\partial x^{i}_{+}} \left( nu^i \Pi^{jk} \right) + nu^i \frac{\partial \Pi^{jk}}{\partial x^{i}_{+}} + n \frac{\partial \Pi^{jk}}{\partial t} \right).
\end{equation}
We then employ
\begin{subequations}
\begin{align}
    \partial _tn & =-\frac{\partial }{\partial x_{+}^i} \left ( nu^i \right ) , \\
    \frac{\partial u^j}{\partial t} +u^i\frac{\partial u^j}{\partial x_{+}^{i}} & =\frac{1}{mn} \frac{\partial \mathcal{C}^{ij} }{\partial x_{+}^{i}} -\frac{1}{m}\frac{\partial V}{\partial x_{+}^{i}} -2\gamma u^j , \\
    \frac{\partial u^k}{\partial t} +u^i\frac{\partial u^k}{\partial x_{+}^{i}} & =\frac{1}{mn} \frac{\partial \mathcal{C}^{ik}}{\partial x_{+}^{i}} -\frac{1}{m}\frac{\partial V}{\partial x_{+}^{k}} -2\gamma u^k ,
\end{align}
\end{subequations}
and define $B^{ijk}$ via
\begin{equation}
B^{ijk} = Q^{ijk} - m^2u^iu^ju^k - \frac{m}{n}\left ( u^i\mathcal{C}^{jk} + u^j\mathcal{C}^{ik} + u^k\mathcal{C}^{ij} \right ) .
\end{equation}
Substituting these expressions into Eq.~\eqref{eq:evolution of second moment} and truncating the hierarchy of moments by setting $B^{ijk}=0$ leads to an evolution equation for the stress tensor
\begin{equation}
    \frac{\partial \mathcal{C}^{jk}}{\partial t}+u^i\frac{\partial \mathcal{C}^{jk}}{\partial x_{+}^i} =-\mathcal{C}^{jk}\frac{\partial u^i}{\partial x_{+}^i}-\mathcal{C}^{ij}\frac{\partial u^k}{\partial x_{+}^i}
    -\mathcal{C}^{ik}\frac{\partial u^j}{\partial x_{+}^i}  -4\gamma \mathcal{C}^{jk}-4\gamma nk_BT\delta ^{jk}.
\end{equation} 
By focusing on the first few moments of the Wigner function, we derived the hydrodynamic equations describing viscous quantum fluids, which are exactly equivalent to the equations derived from the Caldeira-Leggett model in section \ref{sec: Emergent Viscous Hydrodynamics}.
This further demonstrates that the system's long-time dynamics are equivalent to low-momentum, near-equilibrium dynamics, as is expected for standard hydrodynamics.

\bibliography{references}

\end{document}